\documentclass[journal,comsoc]{IEEEtran}
\usepackage{caption}
\usepackage{amsmath, amsfonts, amssymb}
\usepackage{enumitem}
\usepackage{booktabs}
\usepackage{cite}
\usepackage{float}
\usepackage{graphicx}
\usepackage{hyperref}
\usepackage[utf8]{inputenc}
\usepackage{subcaption}
\usepackage{xcolor}
\usepackage[table]{xcolor}

\begin{document}

\title{CUNEC: A Path Loss Model for\\Urban Cell-Free Massive MIMO Networks

\thanks{Part of this work was presented at IEEE Globecom 2022~\cite{choi2022realistic}. T. Choi was, and Y. Zhang, and A. F. Molisch are, with the University of Southern California, Los Angeles, CA, USA. I. Kanno and M. Ito are with KDDI Research, Inc., Saitama, Japan. Email: \{choit, yzhang26, molisch\}@usc.edu.
The work of T. Choi, Y. Zhang, and A. F. Molisch was financially supported by KDDI Research, Inc. and the National Science Foundation (ECCS-1731694).
}
}

\author{\IEEEauthorblockN{Thomas Choi, Yuning Zhang, Issei Kanno, Masaaki Ito, and Andreas F. Molisch}}

\maketitle

\begin{abstract}
Accurate path loss (PL) modeling is essential for evaluating and optimizing cell-free massive MIMO systems, especially in dense urban environments where traditional models fail to capture the complexity of real-world propagation.
This paper introduces CUNEC (\underline{C}ell-free massive MIMO for \underline{U}rban \underline{N}on-stationary \underline{E}nvironments with \underline{C}orrelations), a novel PL model that accounts for spatial non-stationarity, inter-access point (AP)/user equipment (UE) correlations, and urban-specific propagation phenomena such as corner diffraction and street canyon waveguiding.
CUNEC segments AP-UE paths by street order, models PL as a stochastic function of urban geometry, and integrates spatially correlated shadowing.
The parameters are derived from large-scale ray tracing and validated against both additional ray tracing in New York, NY and real-world channel measurements in Los Angeles, CA. 
Compared to the conventional $\alpha$--$\beta$ model, CUNEC significantly improves accuracy in the considered urban propagation scenarios.
An open-source dataset comprising over 30,000 AP locations and 128 UE positions is also released to support reproducible research and future system development.
\end{abstract}

\section{Introduction} \label{sec:introduction}
    \subsection{Motivation}
        Cell-free massive multiple-input multiple-output (CF-mMIMO)\footnote{CF-mMIMO builds upon the related concepts of base station cooperation, distributed MIMO, network MIMO, Coordinated Multipoint (CoMP), and Cloud-Radio Access Network (C-RAN) architectures.} is rapidly emerging as a pivotal technology to meet the ever-increasing demands for high capacity and ubiquitous connectivity in modern wireless networks~\cite{ngo2017cell, demir2021foundations}.
        Unlike traditional cellular systems where a single base station (BS) communicates with all user equipments (UEs) in its cell, CF-mMIMO leverages a distributed architecture in which numerous access points (APs) cooperatively serve the UEs.
        This approach offers significant advantages, including improved spatial diversity, enhanced coverage, and reduced inter-cell interference.
        As a result, CF-mMIMO is already being integrated into advanced 5G deployments and is considered a key enabler for future 6G systems~\cite{tataria20216G, molisch2023wireless, chu2025testbed}.

        A crucial aspect of system design and analysis is the accurate modeling of the wireless channel, particularly the path loss (PL) between APs and UEs.
        However, most CF-mMIMO performance studies rely on spatially stationary PL models based solely on Euclidean distance \cite{
        nayebi2018access, zhang2018performance, björnson2020scalable, xiao2022mobility}, often combined with stochastic geometry such as Poisson Point Processes (PPPs)~\cite{
        papazafeiropoulos2020performance}.
        Although convenient for theoretical analysis, this assumption rarely holds in real-world deployment scenarios.

\begin{figure}[t!]
            \centering
            \includegraphics[width=0.85\linewidth]{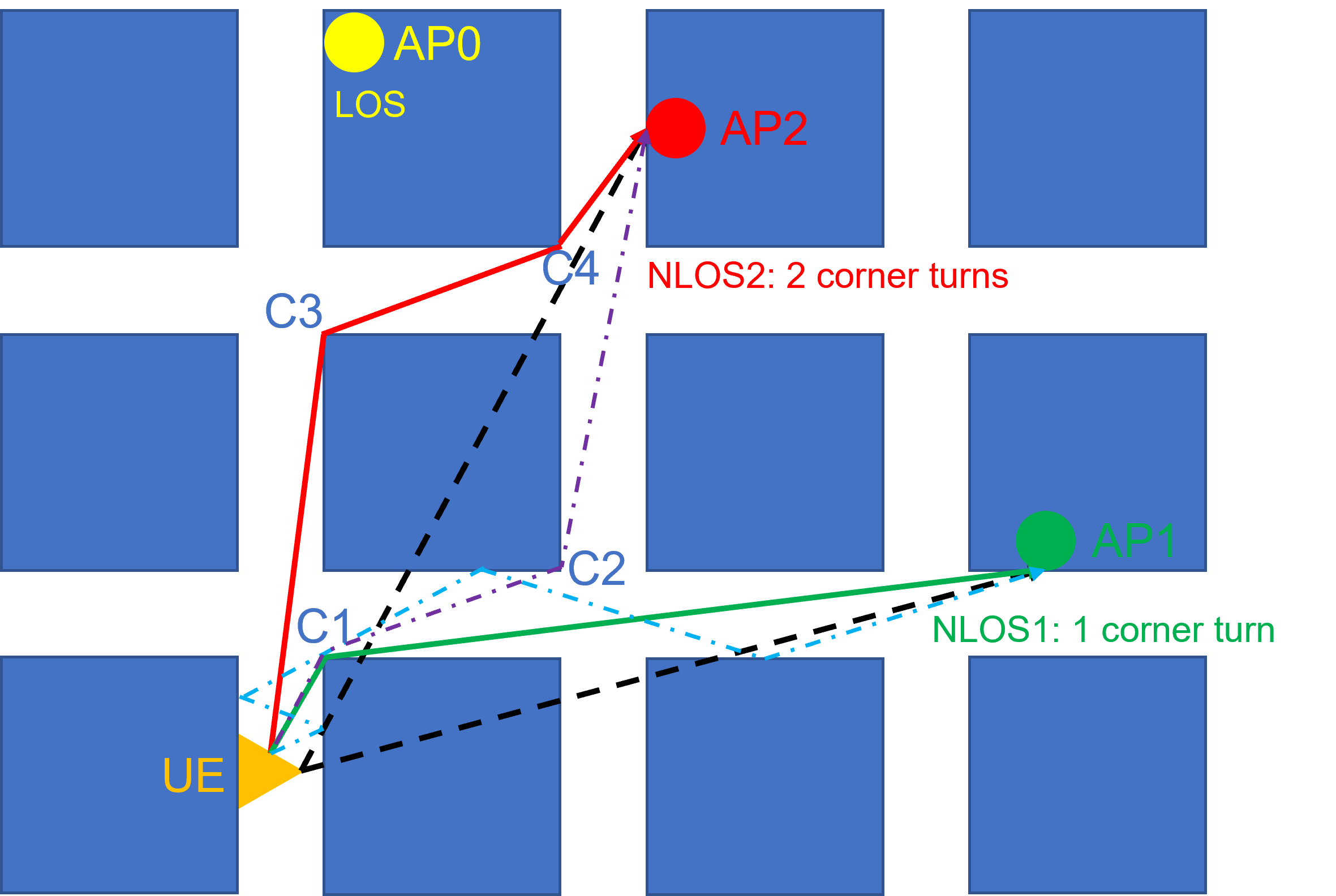}
            \caption{\small Top-down view of multipath propagation in a Manhattan grid, illustrating three equidistant APs serving a single UE.
            Blue squares represent buildings. 
            While both the AP1 and AP2 are at NLOS with the same AP--UE distance, AP2’s paths include large-angle diffractions around two building corners, while AP1’s paths feature a combination of large-angle diffraction around one building corner and small-angle diffraction with multiple reflections, resulting in different PL.}
            \label{fig:motivation}
        \end{figure}

        In particular, most current and anticipated CF-mMIMO deployments are in dense urban environments with antenna heights below rooftops (i.e., microcells). 
        These environments are characterized by high-rise buildings, narrow street canyons, and complex arrangements of scatterers, which in turn lead to phenomena such as corner diffraction, waveguiding along streets, and spatially correlated shadowing.
        Such effects introduce significant deviations from the assumption that PL depends solely on Euclidean distance.
        For example, two APs equidistant from a UE can experience vastly different PL due to the specific arrangement of buildings and obstructions along their respective propagation paths, as illustrated in Fig.~\ref{fig:motivation}. 
        This \emph{spatial non-stationarity} renders traditional Euclidean distance-based models inaccurate and unreliable for predicting the performance of CF-mMIMO systems.
        Furthermore, the numerous APs in a CF-mMIMO network are often densely deployed and located much closer to each other, and this proximity introduces a critical factor: spatial correlation between the large-scale fading of nearby APs.
        This inter-AP correlation fundamentally impacts system-level performance metrics such as user throughput, scheduling, and power control in CF-mMIMO, yet it is often absent from existing models.

        This paper introduces the \textbf{\underline{C}ell-free massive MIMO for \underline{U}rban \underline{N}on-stationary \underline{E}nvironments with \underline{C}orrelations (CUNEC)} model---a novel PL framework designed to address these critical shortcomings.
        CUNEC integrates the fundamental physics and geometric characteristics of radio wave propagation in urban environments, accounting for higher-order non-line-of-sight (NLOS) conditions, corner diffraction, waveguiding effects, and spatially correlated shadowing.
        By parameterizing CUNEC using extensive ray tracing simulations, we achieve a significantly more accurate and realistic representation of the multi-AP, multi-UE channels characteristic of CF-mMIMO in dense urban deployments, as validated by comparisons with independent ray tracing simulations and channel measurements in real-world environments.

    \subsection{Literature survey}
        We categorize the related literature into three key areas: (i) advanced PL modeling techniques that go beyond simple Euclidean distance-based methods, (ii) models specifically tailored for CF-mMIMO system analysis, and (iii) CF-mMIMO measurement campaigns and testbeds.

        \subsubsection{Advanced PL modeling}
            To overcome the limitations of PL models solely based on Euclidean distance, researchers have developed a range of modeling techniques that incorporate more realistic representations of propagation paths and environment-specific variability.
            An early development in this area was the recursive model of~\cite{berg1995recursive}, which introduced the concept of segmenting the propagation path based on turns in an urban grid.
            However, this model assumed uniform parameters across all segments, ignoring street-by-street variability and omitting environmental-specific characteristics.
            Subsequent studies, such as~\cite{inomata2015path, haneda2016frequency, ituP1411}, adopted Manhattan distance metrics and geometric parameters to better align with signal propagation along street canyons, but similarly did not account for PL variations between streets with the same geometric parameters.

            To better capture environmental variability and spatial non-stationarity, studies introduced random variations in the PL coefficient across cells~\cite{erceg1992urban}, between buildings~\cite{ghassemzadeh2004measurement} (for indoor propagation), and streets~\cite{molisch2016spatially, karttunen2017spatially}.\footnote{Different PL coefficients in different streets had been observed experimentally in ~\cite{smith2004urban}}
            However, these works focus on systems with a single BS and are not directly applicable to CF-mMIMO systems.

            More recently, machine learning techniques have shown promise in capturing complex relationships between environmental variables and PL~\cite{zadeh2024machine}.
            However, despite these developments, existing models typically do not capture critical inter-AP correlation effects while simultaneously accounting for detailed urban characteristics such as street canyon geometries, all of which significantly impact the performance of CF-mMIMO systems in realistic deployments.

        \subsubsection{Models for CF-mMIMO system analysis}

            Recent modeling efforts for CF-mMIMO systems have employed various analytical frameworks.
            For example, models based on stochastic geometry, such as~\cite{papazafeiropoulos2020performance}, often assume spatial stationarity by modeling AP and UE locations using homogeneous PPPs, and relying on isotropic PL and fading assumptions as well as homogeneous scattering when applying standard fading distributions throughout the coverage area.
            
            Probabilistic line-of-sight (LOS) models, such as~\cite{mukherjee2022performance}, use distance-dependent LOS probabilities that implicitly assume spatial stationarity if the LOS function depends only on Euclidean distance. 
            These models often use different, but still Euclidean-distance-based pathloss models for LOS and NLOS, as well as different simplified fading distributions (e.g., Ricean for LOS, Rayleigh for NLOS) again under the assumption of spatially homogeneous scattering conditions.
            
            Works that incorporate spatially correlated Ricean~\cite{femenias2020short} or Rayleigh~\cite{wang2022uplink} fading introduce spatial correlation via exponential or jointly correlated models. 
            However, these correlation functions generally depend only on the relative distance between the terminals, which still implies spatial stationarity and may not capture environmental asymmetry or anisotropy arising in urban deployments.
            
            Models based on angle of arrival (AoA), such as the localization method in~\cite{liao2020aoa}, and stochastic models based on standardized geometry such as COST 2100~\cite{liu2012cost}, provide a richer spatial structure by introducing angular clusters and visibility regions. 
            However, while those models might give rise to some non-stationarities, they do not account for the urban waveguiding effects and associated non-stationarity and anisotropic behavior.
            
            In summary, while these models offer valuable tools for CF-mMIMO analysis, most rely on simplifying assumptions such as spatial stationarity, isotropy, and homogeneous scattering. 
            These assumptions can limit their ability to accurately capture the effects of inter-AP correlation and street-level environmental characteristics in realistic urban deployments.

        \subsubsection{CF-mMIMO measurements and prototypes}
            Numerous experimental studies, testbeds, and prototypes have been developed for CF-mMIMO; however, few specifically focus on PL modeling in urban environments.
            Recent efforts include prototype evaluations focusing on calibration, phase synchronization, and channel reciprocity~\cite{cao2023experimental}; testbeds investigating scalability \cite{chu2025testbed} and distributed channel behavior such as channel hardening and channel aging~\cite{Löschenbrand2022towards}; and extensive measurement datasets such as our own ~\cite{choi2022using, zhang2025cell}.
            Although these studies offer valuable empirical insights, they typically do not provide statistical models for the CF-mMIMO channels.

    \subsection{Contributions}
        To overcome these gaps in the literature, the CUNEC model presented here captures spatial non-stationarity, inter-AP and inter-UE PL correlations, and urban-specific propagation mechanisms---such as higher-order NLOS conditions. 
        Specifically, this paper makes the following key contributions:
        \begin{enumerate}[leftmargin=*]
            \item{\textbf{Development of the CUNEC model}}: We propose a PL modeling approach that explicitly accounts for corner diffraction, waveguiding, and other urban-specific propagation effects, while incorporating variability in PL and its parameters across different streets.
            \item{\textbf{Advanced parameterization using extensive ray tracing}}: Through large-scale ray tracing simulations covering AP--UE distances up to 500\,m, street widths ranging from 15\,m to 40\,m, and building heights from 20\,m to 45\,m, we derive key model parameters applicable to a wide range of urban layouts.
            \item{\textbf{Empirical validation}}: CUNEC is validated using high-fidelity ray tracing simulations in an urban environment in New York, NY, USA and real-world channel measurements from the University of Southern California (USC) campus~\cite{choi2021energy}, Los Angeles, CA, USA, demonstrating substantial accuracy improvements over the conventional $\alpha$--$\beta$ model.
            \item{\textbf{Open-access measurement dataset}}: To encourage further research, we release a comprehensive CF-mMIMO dataset comprising more than 30,000 AP locations, 128 UE positions, and 2,301 frequency points, along with detailed metadata on AP--UE geometries, propagation characteristics, and environmental parameters.
        \end{enumerate}


\section{Alpha-Beta Path Loss Modeling} \label{sec:conv_model}
    \subsection{Path loss definitions and large-scale fading}
        In this section, we briefly review the traditional \(\alpha\)--\(\beta\) PL model. 
        For notational convenience, we assume a downlink scenario---i.e., the AP as the transmitter (TX) and the UE as receiver (RX)---though it is worth noting that PL is reciprocal. 
        The PL is defined as:
        \begin{equation} \label{eq:total}
            PL^{\mathrm{lin}}(\mathbf{r}_{\mathrm{AP}}, \mathbf{r}_{\mathrm{UE}}) = \mathbb{E}_{\mathrm{ssf}}\Bigl\{\tfrac{P_{\mathrm{AP}}}{P_{\mathrm{UE}}}\Bigr\},
        \end{equation}
        where \(\mathbf{r}_{\mathrm{AP}}\) and \(\mathbf{r}_{\mathrm{UE}}\) denote the locations of the AP and UE, respectively, and \(P_{\mathrm{AP}}\) and \(P_{\mathrm{UE}}\) are transmit and receive powers.
        The operator \(\mathbb{E}_{\mathrm{ssf}}\{\cdot\}\) denotes the expectation over small-scale fading in linear scale. Throughout this paper, $PL^\mathrm{lin}$ denotes PL expressed in linear scale, while $PL$ denotes PL in dB scale.
        The path gain is $PG^{\mathrm{lin}}\!=\!1/PL^{\mathrm{lin}}$ and thus $PG\!=\!-PL$.

        PL can be decomposed into two components:
        \begin{equation}\label{eq:full_PL}
            PL(\mathbf{r}_{\mathrm{AP}}, \mathbf{r}_{\mathrm{UE}}) 
            = \overline{PL}(\mathbf{r}_{\mathrm{AP}}, \mathbf{r}_{\mathrm{UE}}) 
            + S(\mathbf{r}_{\mathrm{AP}}, \mathbf{r}_{\mathrm{UE}}),
        \end{equation}
        where \(S\) accounts for large-scale fading (also called shadowing), related to environmental variability~\cite[Chapter 7]{molisch2023wireless}, while \(\overline{PL}\) denotes the PL averaged over the large-scale fading.
        In many models, \(S\) is treated as a zero-mean Gaussian random variable with standard deviation \(\sigma_S\).

    \subsection{The Alpha-Beta model and its shortcomings}
        A widely used approach to PL modeling is the \(\alpha\)--\(\beta\) framework, which forms the basis of empirical methods such as the Okumura-Hata model~\cite{hata1980empirical} and has influenced standardization efforts in 3GPP~\cite{3gpp}.
        In this model, \(\overline{PL}_{\alpha\text{-}\beta}\) without the shadowing term from~\eqref{eq:full_PL} is modeled as a deterministic function of the Euclidean distance between AP and UE, $d=||\mathbf{r}_{\mathrm{AP}}-\mathbf{r}_{\mathrm{UE}}||_2$, where $||.||_2$ denotes the L2 norm:
        \begin{align}
            \overline{PL}_{\alpha\text{-}\beta}(d) 
            &= \alpha + 10\,\beta\,\log_{10}(d/d_\mathrm{ref}) \notag \\
            &= \Delta + 10\,\beta\,\log_{10}(d) + \mathrm{FSPL}_{1\,\mathrm{m}},
        \end{align}
        where \(\alpha\) is the reference PL at distance \(d_\mathrm{ref}\), \(\beta\) is the  PL exponent, \(\mathrm{FSPL}_{1\,\mathrm{m}}\) denotes the free-space PL at 1\,m, and \(\Delta\! =\! \alpha\! -\! \mathrm{FSPL}_{1\!\,\mathrm{m}}\) accounts for the environment-dependent offset relative to the free-space reference when \(d_\mathrm{ref}\!=\!1\)\,m.
        This model is often combined with specific assumptions about the large-scale fading---namely that \(\sigma_S\) is independent of $\mathbf{r}_{\mathrm{AP}}$ and $\mathbf{r}_{\mathrm{UE}}$, and that the normalized autocorrelation function $\rho_{SS}$ depends only on \(d\).
        In particular, the Gudmundson model \(\rho_{SS}(d)\!=\!\exp\!\left(-d/d_{\mathrm{corr}}\right)\), where \(d_{\mathrm{corr}}\) is the correlation distance, is widely used \cite{gudmundson1991correlation}.

        Despite its widespread use and historical significance, the \(\alpha\)--\(\beta\) model exhibits key limitations when applied to urban CF-mMIMO systems:
        \begin{itemize}[leftmargin=*]
            \item \textbf{Dependence on Euclidean distance}: While the \(\alpha\)--\(\beta\) model shows dependence solely on the Euclidean distance, propagation in urban microcellular deployments is influenced by the individual path segments along street canyons and the number of corner diffractions (``bending around the corners"). 
            Consequently, the average PL model should depend on both \(\mathbf{r}_{\mathrm{AP}}\) and \(\mathbf{r}_{\mathrm{UE}}\), as well as the surrounding environment, rather than solely on \(d\).
            \item \textbf{Assumption of stationarity}: The model assumes that the parameters of both the average PL and the shadowing are constant across the entire radio environment.
            However, both measurements and ray tracing studies have shown that this assumption does not hold in urban environments, particularly in microcellular deployments.
            In reality, the parameters vary from street to street and should be treated as random variables with their own distributions.
            \item \textbf{Lack of inter-AP correlation}: While standard \(\alpha\)--\(\beta\) models account for correlation between UEs as a function of their separation, they typically ignore inter-AP correlations or approximate them using fixed correlation coefficients.
            Such approaches neglect the influence of the actual spatial configuration of the APs, which can significantly impact performance.
        \end{itemize}


\section{The CUNEC Model} \label{sec:CUNEC}
    The CUNEC model is a channel modeling framework specifically developed for urban CF-mMIMO environments.
    This approach addresses a key shortcoming of the conventional \(\alpha\)--\(\beta\) PL model---namely, its assumption that PL depends solely on Euclidean distance, an assumption that breaks down in urban microcellular environments--- by employing a segmented-path methodology that decomposes the propagation path between the AP and UE into distinct street segments, and accounts for the ``bending around the corner".
    The total PL is then computed as the cumulative sum of the PL contributions from each individual segment.
    
    Furthermore, CUNEC captures environmental variability by modeling PL coefficients and shadowing variances as random variables that vary from street to street, reflecting the heterogeneous nature of dense urban environments and eliminating the oversimplified assumption of parameter stationarity.
    Additionally, CUNEC explicitly models spatial correlation effects among multiple APs and UEs, as well as localized propagation mechanisms that dominate in dense urban deployments.

    The remainder of this section elaborates on CUNEC’s core components: the street-order segmentation approach, combining multiple dominant paths, inter-AP and inter-UE shadowing correlations, and the treatment of over-the-rooftop (ORT) propagation in dense urban deployments.

    \subsection{CUNEC at different street orders} \label{sec:street_order}
        The first step in the CUNEC model is to identify feasible propagation paths between APs and UEs based on the minimum number of corner turns---referred to as the street order.
        As an example, consider the propagation between the UE and AP1 in Fig.~\ref{fig:motivation}.
        The LOS path is obstructed, but a viable connection is established via a single corner turn at point C1.
        In contrast, propagation from the UE to AP2 requires two turns around building corners, at C3 and C4.

        We can thus define the following types of paths\footnote{Note that a ``path" in this context is {\em not} a multipath component, but rather a general route by which signals can propagate from TX to RX.} between an AP and UE:
        \begin{itemize}[leftmargin=*]
            \item \textbf{Zeroth-order street (LOS)}: The AP and UE lie on the same street segment, with minimal or no corner diffraction.
            \item \textbf{First-order street (NLOS1)}: The path involves a single corner, introducing some diffraction and reflection losses.
            \item \textbf{Second-order street (NLOS2)}: The path involves two corners, resulting in higher diffraction and reflection losses.
        \end{itemize}
        Higher-order streets (i.e., routes with three or more corner turns) are omitted because in a predominantly rectangular city-grid, alternative one- or two-corner paths typically exist and dominate propagation. Paths with additional turns experience substantially higher attenuation and thus are expected to contribute negligibly to the overall path loss, especially in dense AP deployments.

        For the PL modeling, CUNEC consider only the lowest-order paths between an AP and UE, as these typically dominate due to lower attenuation.
        For example, if a LOS path exists, higher-order alternatives such as NLOS1 and NLOS2 paths are disregarded, even if they provide valid propagation paths, since each additional corner turn generally incurs significant excess loss.
        However, if multiple paths of the same order exist (e.g., the red and the purple path in Fig.~\ref{fig:motivation}), all are considered, as discussed in Sec.~\ref{sec:multi_path_combine}.

        The propagation path is then partitioned into segments consisting of straight line components.
        A LOS path consists of a single segment.
        An NLOS1 path includes two segments: one from the AP to the first corner and another from the corner to the UE.
        Similarly, an NLOS2 path consists of three segments.
        CUNEC assumes that each segment incurs its own propagation loss, thereby capturing street-specific phenomena such as diffraction and corner losses.
        The total PL is computed as the sum of the contributions from individual street segments:
        \begin{equation} \label{eq:total_PL}
            PL(\mathbf{r}_{\mathrm{AP}}, \mathbf{r}_{\mathrm{UE}}) 
            = \sum_{n=0}^{N} \overline{PL}_n(\mathbf{r}_{\mathrm{AP}}, \mathbf{r}_{\mathrm{UE}}) + \mathrm{S}_N.
        \end{equation}
        where \(\overline{PL}_n(\mathbf{r}_{\mathrm{AP}}, \mathbf{r}_{\mathrm{UE}})\) denotes the average PL contribution from the \(n\)-th street segment, \(N\) is the street order, and \(S_N\) denotes the shadowing component associated with an \(N\)-th street link.
        The functional forms used to calculate \(\overline{PL}_n\) and the methodology for generating \(\mathrm{S}_N\) (using $\sigma_{\mathrm{S\text{-}}N}$ and $\mathrm{d}_{\mathrm{corr\text{-}}N}$) are detailed below. 

        We emphasize that CUNEC model does not represent exact electromagnetic solutions, but are instead heuristic fitting functions whose form is guided by the analysis of extensive ray tracing simulations (see Sec.~\ref{sec:rect}).
        Among other limitations, we observe that the functional dependence on UE movement differs from that of AP movement.\footnote{However, due to reciprocity, uplink and downlink PL are always identical.}
        While the results align well with the scenarios we investigated, it can be observed that in the limiting case where the AP and UE are at the same height (i.e., a true device-to-device scenario), a different functional form with explicit symmetry would be more appropriate.

        \subsubsection{Zeroth-order streets (LOS)} \label{sec:model_zeroth}
            In zeroth-order streets, where Manhattan and Euclidean distances coincide, the average PL is expressed as:
            \begin{equation} \label{eq:PL0}
                \overline{PL}_{0}(\mathbf{r}_{\mathrm{AP}}, \mathbf{r}_{\mathrm{UE}})
                = \Delta_{0} + 10 \,\mathrm{b}_{0}\,\log_{10}(d_0) + \mathrm{FSPL}_{\mathrm{1m}},
            \end{equation}
            where \(d_0\!=\!\bigl\|\mathbf{r}_{\mathrm{AP}}- \mathbf{r}_{\mathrm{UE}}\bigr\|_2\) is the distance between the AP and UE along the zeroth-street, \(\Delta_0\) is an offset capturing deviations from the free-space reference, and \(\mathrm{b}_0\) is the PL exponent.
            
            In contrast to \(\Delta\), \(\beta\), \(\sigma_S\), and \(d_{\mathrm{corr}}\) in the conventional \(\alpha\)--\(\beta\) framework---typically treated as fixed parameters---CUNEC models \(\Delta_0\), \(\mathrm{b}_0\), \(\sigma_{\mathrm{S}\text{-}0}\), and \(\mathrm{d}_{\mathrm{corr}\text{-}0}\) as \emph{random variables} to capture local variations in building materials, street widths, and other environment-specific factors.

        \subsubsection{First-order streets (NLOS1)} \label{sec:model_first}
            First-order streets involve a single corner turn along the propagation path.
            This path is naturally divided into two segments: (i) from the AP to the corner, and (ii) from the corner to the UE.
            
            We define \(\mathbf{r}_{\mathrm{c\text{-}1}}\) as the location of the corner in the first-order street, \(d_{\mathrm{c}}\!=\!\bigl\|\mathbf{r}_{\mathrm{AP}} - \mathbf{r}_{\mathrm{c\text{-}1}}\bigr\|_2\) as the distance from the AP to the corner, and \(d_1\!=\!\bigl\|\mathbf{r}_{\mathrm{c\text{-}1}} - \mathbf{r}_{\mathrm{UE}}\bigr\|_2\) as the distance from the corner to the UE.
            The PL for the AP-to-corner segment follows the same model used for zeroth-order streets.
            The average PL for the corner-to-UE segment is then modeled as:
            \begin{align} \label{eq:PL1}
                \overline{PL}_{1}&(\mathbf{r}_{\mathrm{AP}}, \mathbf{r}_{\mathrm{UE}})
                =\; \left[\Delta_{1} + C \cdot \mathbb{I}(\mathbf{r}_{\mathrm{AP}}, \mathbf{r}_{\mathrm{UE}}, \mathbf{r}_{\mathrm{c\text{-}1}})
                \right] \cdot \left(1 - e^{-\kappa d_1}\right) \nonumber\\[1ex]
                & + \left[\frac{1}{\pi} \arctan\left(\frac{1}{100} (d_c - d_{\mathrm{th}})\right) + \frac{1}{2} \right] \cdot \left(10\, \mathrm{b}_1 \log_{10}(d_1)\right)
            \end{align}
            In this expression:
            \begin{itemize}
                \item \(\Delta_{1}\) is the base corner loss (in dB), representing the fundamental diffraction loss component that increases as \(d_1\) increases.\footnote{Note that our model for the corner loss is a heuristic, not directly related to the fundamental diffraction laws. This is motivated by the fact that we have diffractions from corners of the same intersection, that add up in a way that is hard to describe with analytical equations.}
                \item \(C\) is an offset parameter (in dB) that accounts for additional loss when the AP or the UE is positioned away from the walls that form the corner of diffraction. 
                In such geometries, the corner loss tends to be higher because of the reduced interaction with the corner, as observed in our simulations.
                \item Under the assumption that the AP is in a horizontal street and the UE is in a vertical street or vice-versa, we define the indicator function \(\mathbb{I}(\mathbf{r}_{\mathrm{AP}}, \mathbf{r}_{\mathrm{UE}}, \mathbf{r}_{\mathrm{c\text{-}1}})\) as:
                \[
                \mathbb{I}=
                \begin{cases}
                1, & \text{if } 
                \min\left(|x_{\mathrm{AP}} - x_{\mathrm{c1}}|, |x_{\mathrm{UE}} - x_{\mathrm{c1}}|\right) > w/2 \\[0.5ex]
                & \quad \text{or } 
                \min\left(|y_{\mathrm{AP}} - y_{\mathrm{c1}}|, |y_{\mathrm{UE}} - y_{\mathrm{c1}}|\right) > w/2 \\[0.5ex]
                0, & \text{otherwise}
                \end{cases}
                \]
                Here, \(\mathbf{r}_{\mathrm{AP}} = [x_{\mathrm{AP}}, y_{\mathrm{AP}}]^T\), \(\mathbf{r}_{\mathrm{UE}} = [x_{\mathrm{UE}}, y_{\mathrm{UE}}]^T\), and \(\mathbf{r}_{\mathrm{c\text{-}1}} = [x_c, y_c]^T\) are the coordinates of the AP, UE, and corner, respectively. 
                \(w\) denotes the width of the street. The function returns 1 if either the AP or UE is located further than half the street width from the walls containing the corner.
                \item \(\kappa\) controls the rate at which the corner loss term, \(\Delta_{1}\!\cdot\!(1\!-\!e^{-\kappa\cdot d_1})\), transitions from low to high values as \(d_1\) increases.\footnote{While millimeter-wave (mmWave) models~\cite{karttunen2017spatially} may use a step function to represent corner loss, a gradual transition is more appropriate at lower frequencies, where MPCs can bend around obstacles.}
                \item \(10\,\mathrm{b}_1\,\log_{10} (d_1)\) represents the standard distance-dependent loss for the segment after the corner, where \(\mathrm{b}_1\) is a stochastic PL exponent that captures environmental variability. This term is included only for $d_1>1$m.
                \item The bracketed term, \(\frac{1}{\pi}\arctan\Bigl(\!10\bigl(d_c - d_{\mathrm{th}}\bigr)\!\Bigr)+\frac{1}{2},\) is a smooth activation function that transitions from 0 to 1 around the \emph{diffraction threshold distance} \(d_{\mathrm{th}}\). 
                For \(d_\mathrm{c}\) well below \(d_{\mathrm{th}}\), this factor is close to 0, diminishing the contribution of the distance-dependent loss.
                Conversely, for \(d_\mathrm{c}\) well above \(d_{\mathrm{th}}\), it approaches 1, fully applying the distance-dependent loss.
                This formulation is motivated by the observation that, until \(d_\mathrm{c}\) exceeds \(d_{\mathrm{th}}\), the dominant MPC is typically a reflection rather than a diffracted path.\footnote{Throughout this paper, we set \(d_\mathrm{th}\!=\!70\)\,m based on extensive simulations.}
            \end{itemize}

            As in the zeroth-order model, CUNEC models \(\Delta_{1}\), \(\mathrm{b}_1\), \(C\), \(\kappa\), \(\sigma_{\mathrm{S}\text{-}1}\), and \(\mathrm{d}_{\mathrm{corr}\text{-}1}\) as correlated random variables.
            This stochastic approach captures both the inherent variability of corner diffraction and reflection phenomena, as well as the complex interdependencies between environmental factors and model parameters in urban NLOS scenarios.

        \subsubsection{Second-order streets (NLOS2)}
            For second-order streets, the signal undergoes two corner turns, typically placing the AP and UE on parallel streets.
            Using \(\mathbf{r}_{\rm c\text{-}2}\) to denote the location of the second corner-around which the signal must diffract to reach the UE and \(d_2\!=\! \|\mathbf{r}_{\mathrm{c\text{-}2}} - \mathbf{r}_{\mathrm{UE}}\|_2\), the average PL for the segment after the second corner turn is expressed as:
            \begin{equation} \label{eq:PL2}
                \overline{PL}_{2}(\mathbf{r}_{\rm AP},\mathbf{r}_{\rm UE})
                = 10\,\mathrm{b}_{2}\,\log_{10}(d_2),
            \end{equation}
            where \(\mathrm{b}_{2}\) is the stochastic PL exponent for the post-corner segment.
            
            Ray tracing simulations indicate that second-order corner losses do not exhibit saturation behavior of corner loss, thereby eliminating the need for a logistic or activation term in this segment.
            Moreover, the high overall PL—--especially under waveguiding conditions—--renders second-order (and higher-order) paths impractical for dense urban deployments, which justifies the simplified formulation in~\eqref{eq:PL2}.

    \subsection{Combining multiple dominant paths} \label{sec:multi_path_combine}
        While the preceding subsections assume a single dominant route per street order, real urban environments may give rise to multiple paths of comparable strength---as illustrated by the case of AP2 in Fig.~\ref{fig:motivation}.
        In such cases, CUNEC combines the PLs by summing the corresponding {\em linear PGs}, thereby accounting for the independent contributions of each route’s power to the total received power.

        Thus, let \(\{PL^\mathrm{lin}_{n\text{-}i}\}_{i=1}^{K}\) denote the linear-scale PLs of \(K\) dominant paths at the \(n\)-th street order.
        The combined linear-scale PL, \(PL^\mathrm{lin}_n\), is then given by:
        \begin{equation}
            \frac{1}{PL^\mathrm{lin}_n} = \sum_{i=1}^{K} \frac{1}{PL^\mathrm{lin}_{n\text{-}i}}.
        \end{equation}
        This approach assumes that each path contributes additively in the power domain, consistent with the assumption that small-scale fading has been averaged out.\footnote{This assumption is justified because, in practice, different propagation paths between two points (e.g., going ``top–right'' or ``right–top'' around a building block) arrive with independent phases due to differences in geometry and scattering.
        The rapid constructive and destructive interference terms cancel, and the expected received power equals the sum of the average powers of the individual paths.}

        However, we note that the component PLs along different routes may still exhibit correlation.
        For example, the NLOS2 PLs from AP to corner C2 (part of path 1) and from AP to corner C3 (part of path 2) both traverse the same vertical street segment in Fig.~\ref{fig:motivation}.
        Thus, the PLs for these constituent segments are LOS street segments with the same starting point but different end points, propagating along the same street.
        As a result, they share the same (randomly generated) $\mathrm{b}_0$ and $\Delta_0$, and exhibit correlated shadowing.

    \subsection{Generating correlated shadowing for multi-AP / multi-UE}
        To derive the term $\mathrm{S}_N$ from \eqref{eq:total_PL} for spatially correlated AP locations, we analyzed the shadowing correlations among multiple APs serving a single UE along a single street in the conference version of this paper~\cite{choi2022realistic}.
        We now extend the analysis to scenarios with multiple UEs, or a UE at different locations along a trajectory,\footnote{Regarding temporal correlation (mobility), it can naturally be interpreted as the correlation between two UE positions, reflecting temporal changes due to movement of the UE. We emphasize that we are not making any claims with respect to moving scatterers or blockers.} because accurately capturing shadowing correlation across both APs and UEs is essential for realistic CF-mMIMO performance evaluation.

        We model the large-scale shadowing field \(\mathbf{S}_{\mathrm{gen}} \in \mathbb{R}^{M \times N}\) between \(M\) UEs and \(N\) APs using a two-dimensional anisotropic Gaussian process. 
        The spatial correlation between shadowing values is governed by differing correlation distances along the UE and AP axes.
           
        To generate a realistic realization \(\mathbf{S}_{\mathrm{gen}}\), we follow four main steps:

        \subsubsection{Obtain standard deviations and correlation distances}  
        We begin with shadowing matrices from ray tracing, which are obtained by subtracting $\overline{PL}$ using equations from Sec.~\ref{sec:street_order}:
        \begin{itemize}
            \item \(\mathbf{S}_{\mathrm{UE}} \in \mathbb{R}^{M \times N_0}\), measured across \(M\) UEs for \(N_0\) fixed APs ($N_0 < M$),
            \item \(\mathbf{S}_{\mathrm{AP}} \in \mathbb{R}^{M_0 \times N}\), measured across \(N\) APs for \(M_0\) fixed UEs ($M_0 < N)$.
        \end{itemize}
        From these, we compute \(N_0\) instances of \(\sigma_{\mathrm{S}\text{-}\mathrm{UE}}\) and \(\mathrm{d}_{\mathrm{corr,UE}}\) from \(\mathbf{S}_{\mathrm{UE}}\) and \(M_0\) instances of \(\sigma_{\mathrm{S}\text{-}\mathrm{AP}}\) and \(\mathrm{d}_{\mathrm{corr,AP}}\) from \(\mathbf{S}_{\mathrm{AP}}\).

        \subsubsection{Construct the joint covariance model}  
        We define the covariance between shadowing values \(S(i,j)\) and \(S(k,l)\), for UE indices \(i,k\in\{1,\ldots,M\}\) and AP indices \(j,l\in\{1,\ldots,N\}\), using an anisotropic exponential model:
        \begin{align}
        \Sigma_{(i,j),(k,l)} = \exp\Bigg( -\Bigg[
        \left( \frac{\|\mathbf{r}_{\mathrm{UE},i} - \mathbf{r}_{\mathrm{UE},k}\|}{\mathrm{d}_{\mathrm{corr,UE}}} \right)^2 
        + \left( \frac{\|\mathbf{r}_{\mathrm{AP},j} - \mathbf{r}_{\mathrm{AP},l}\|}{\mathrm{d}_{\mathrm{corr,AP}}} \right)^2 
        \Bigg]^{1/2} \Bigg)
        \end{align}
        where \(\mathbf{r}_{\mathrm{UE},i}\) and \(\mathbf{r}_{\mathrm{AP},j}\) denote the positions of the \(i\)-th UE and \(j\)-th AP, respectively. 
        This defines a \((MN) \times (MN)\) covariance matrix \(\boldsymbol{\Sigma}_{\mathrm{joint}}\) over all UE–AP links.

        \subsubsection{Sample and scale}
        We draw a realization \(\tilde{\mathbf{S}}_{\mathrm{gen}} \in \mathbb{R}^{M \times N}\) from the zero-mean multivariate Gaussian distribution:
        \begin{equation}
        \mathrm{vec}(\tilde{\mathbf{S}}_{\mathrm{gen}}) \sim \mathcal{N}(\mathbf{0}, \boldsymbol{\Sigma}_{\mathrm{joint}}),
        \end{equation}
        and reshape it into matrix form. Since \(\tilde{\mathbf{S}}_{\mathrm{gen}}\) is unscaled with unit variance, we apply the following direct scaling to obtain \(\mathbf{S}_{\mathrm{gen}}\):
        \begin{equation}
        \mathbf{S}_{\mathrm{gen}} = \mathbf{D}_{\mathrm{UE}} \cdot \tilde{\mathbf{S}}_{\mathrm{gen}} \cdot \mathbf{D}_{\mathrm{AP}},
        \end{equation}
        where \(\mathbf{D}_{\mathrm{UE}} \in \mathbb{R}^{M \times M}\) is a diagonal matrix with entries:
        \[
        \left[ \mathbf{D}_{\mathrm{UE}} \right]_{i,i} = \frac{\mathop{\mathbb{E}}\{\sigma_{\mathrm{S\text{-}UE}}\}}{\mathrm{std}(\tilde{\mathbf{S}}_{\mathrm{gen}}(i,:))},
        \]
        and \(\mathbf{D}_{\mathrm{AP}} \in \mathbb{R}^{N \times N}\) is a diagonal matrix with entries:
        \[
        \left[ \mathbf{D}_{\mathrm{AP}} \right]_{j,j} = \frac{\mathop{\mathbb{E}}\{\sigma_{\mathrm{S\text{-}AP}}\}}{\mathrm{std}(\tilde{\mathbf{S}}_{\mathrm{gen}}(:,j))}.
        \]

        \subsubsection{Validate the methodology}
        To validate the proposed shadowing generation framework, we compare the synthetic matrix \(\mathbf{S}_{\mathrm{gen}}\) with directly ray-traced shadowing \(\mathbf{S}_{\mathrm{meas}}\) from a 100\,UE \(\times\) 100\,AP NLOS1 scenario involving rectangular buildings given in Sec \ref{sec:rect}.\footnote{We note that the validation has been conducted primarily on ray tracing datasets due to the absence of measurement data involving both TX and RX mobility.} The synthetic \(\mathbf{S}_{\mathrm{gen}}\) was constructed using statistics obtained from two reduced ray-traced simulations: (i) a UE-side shadowing matrix \(\mathbf{S}_{\mathrm{UE}} \in \mathbb{R}^{100 \times 4}\) with fixed APs, and (ii) an AP-side matrix \(\mathbf{S}_{\mathrm{AP}} \in \mathbb{R}^{4 \times 100}\) with fixed UEs.
        
        Table~\ref{tab:validation} shows the comparison across key statistical metrics, including total standard deviation, per-UE and per-AP variability, and estimated correlation distances. The results confirm that the generated \(\mathbf{S}_{\mathrm{gen}}\) closely reproduces the key statistical properties of the full measured dataset, despite being derived from significantly smaller input dimensions.

        The above considerations deal with the case that the two considered APs are in the same street, and the two UEs are in the same street as well (though the AP street and the UE street can of course be different). If APs are in different streets but UEs in the same street, then a standard one-dimensional shadowing process with exponential correlation for the links from the UEs to each AP holds, while the links to the two different APs are uncorrelated; a similar case holds for APs in the same street but UEs in different streets. When APs and UEs are both in different streets, then all links are uncorrelated. The uncorrelatedness of links starting/ending in different streets was verified from the ray tracing, with correlation coefficients generally below 0.1. 

        \renewcommand{\arraystretch}{1.2}        
        \begin{table}[t]
        \centering
        \caption{Validation of shadowing generation: comparison of measured and generated statistics (\(100 \times 100\) NLOS1 case)}
        \label{tab:validation}
        \begin{tabular}{|l|c|c|}
        \hline
        \textbf{Metric} & \textbf{\(S_{\mathrm{meas}}\)} & \textbf{\(S_{\mathrm{gen}}\)} \\
        \hline
        $\sigma_\mathrm{S}$ (across all elements) & 6.82\,dB & 6.65\,dB \\
        Mean of $\sigma_\mathrm{S\text{-}UE}$ (row) & 6.49\,dB & 6.51\,dB \\
        Mean of $\sigma_\mathrm{S\text{-}AP}$ (col) & 5.34\,dB & 5.34\,dB \\
        Mean of $\mathrm{d}_\mathrm{corr\text{-}UE}$ (row) & 15.41\,m & 17.2\,m \\
        Mean of $\mathrm{d}_\mathrm{corr\text{-}AP}$ (col) & 5.17\,m & 5.43\,m \\
        \hline
        \end{tabular}
        \end{table}

    \subsection{Over-the-rooftop propagation and other effects} \label{sec:otr}
        Over-the-rooftop (ORT) propagation is excluded from the current CUNEC model due to its limited impact in dense metropolitan scenarios, where tall buildings and sub-rooftop AP placements typically dominate~\cite{shaik2020cell}.
        In such settings, waveguiding within street canyons serves as the dominant propagation mechanism.
        
        For completeness, if ORT were to be incorporated, the total PL could be modeled by combining the ORT and waveguided components in the power domain---analogous to the approach used for aggregating multiple dominant paths. 
        Specifically, the total PL can be expressed as:
        \begin{equation}
            \frac{1}{PL^\mathrm{lin}_\text{tot}}
            = \frac{1}{PL^\mathrm{lin}_\mathrm{ORT}} + \frac{1}{PL^{\mathrm{lin}}},
        \end{equation}
        where \(PL^\mathrm{lin}_\mathrm{ORT}\) represents the ORT PL, which can be estimated using established models such as the Walfish-Ikegami model from COST231~\cite{damosso1998digital}, and \(PL^\mathrm{lin}\) denotes the waveguided PL.
        
        Notably, as the street order increases, \(PL^\mathrm{lin}\) typically rises due to compounded diffraction and reflection losses, whereas \(PL^\mathrm{lin}_\mathrm{ORT}\) depends primarily on the Euclidean distance between the AP and UE.
        This divergence suggests that ORT propagation may become more pronounced in higher-order streets.

           
            Finally, we note that foliage and street clutter can influence PL, and their effects may be modeled deterministically as functions of foliage and clutter density, frequency, and additional random variations~\cite{abbasi2024ultra}.
            While CUNEC can be extended to explicitly incorporate these effects into the model parameters, such generalizations are beyond the scope of this paper and are not further considered here.

\section{Ray Tracing and Measurement Scenarios} \label{sec:scenarios}
    Both the \(\alpha\)--\(\beta\) and CUNEC models were parameterized using ray tracing simulations conducted in a regular, rectangular-grid urban environment.
    While we acknowledge that ray tracing cannot capture all aspects of real-world propagation due to its simplified assumptions, this canonical setting provides a controlled and representative geometry for systematic model development. 
    The parameters were then validated using two independent datasets: (i) a detailed ray tracing model of Northwest Manhattan, New York City, NY, USA, and (ii) an outdoor CF-mMIMO measurement campaign conducted on the USC campus. 
    Together, these datasets provide a robust foundation for comparing the \(\alpha\)--\(\beta\) and CUNEC models, as discussed in Sec.~\ref{sec:emp}.

    All ray tracing simulations were conducted using Remcom Inc.’s \textit{Wireless InSite}~\cite{wirelessinsite}, with the following configuration: concrete buildings (the dominant construction material in the Manhattan urban environment), asphalt streets, a carrier frequency of 3.5~GHz, up to 40~MPCs, a maximum of 15 reflections, one diffraction per path, and diffuse scattering disabled---imposed by runtime and storage constraints. Isotropic antennas were selected on both TX and RX.

    \subsection{Rectangular grid of buildings} \label{sec:rect}
\begin{figure}[t!]
            \centering
            \includegraphics[width=0.7\linewidth]{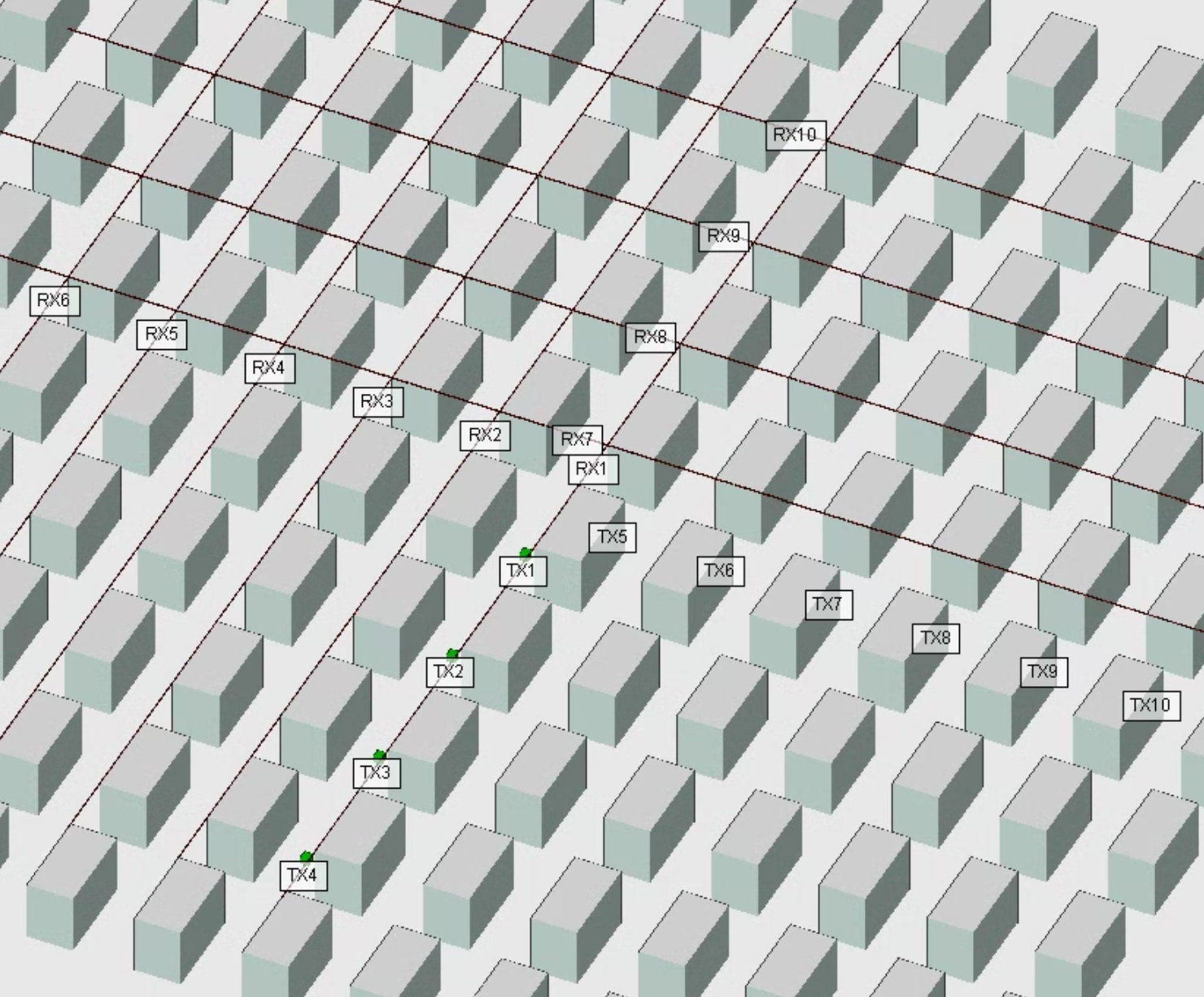}
            \caption{Rectangular-grid environment featuring 20\,m-tall concrete buildings. 
            TX points represent UEs at 1.5\,m height, while RX trajectories correspond to AP trajectories at rooftop level.}
            \label{fig:rt_simple}
        \end{figure}

        Ray tracing simulations were initially performed in a simplified rectangular-grid environment (Fig.~\ref{fig:rt_simple}) to systematically derive model parameters and analyze correlations among them.
        This controlled setting, which mimics typical urban layouts, enabled systematic variation of building dimensions and street widths, allowing us to explore the impact of different street parameters in a structured manner.

        In these simulations:
        \begin{itemize}[leftmargin=*]
            \item Building heights ranged from 20--60\,m; street widths varied between 15--30\,m; and APs were placed 0--5\,m below rooftop height.
            \item Ten UEs were positioned at ground level at 1.5\,m height, while APs were simulated along 8--10 trajectories with densely spaced (with 0.2\,m separation) points, resulting in 80--100 unique parameter sets per scenario. 
            The roles were then reversed---with ten APs and 8--10 UE trajectories. Additional simulations with 100 AP and 100 UE locations were used for confirming the shadowing model as discussed in Sec. III.C, but were not further employed for the model parameterization. 
            \item The maximum AP-UE distance was set to 500\,m.
            While this exceeds the typical communication range between a UE and its serving APs, it is useful for extending the model's validity range-particularly for analyzing interference and optimizing cluster size. 
        \end{itemize}

    \subsection{Northwest Manhattan} \label{sec:manhattan_nw}
\begin{figure}[t!]
            \centering
            \includegraphics[width=\linewidth]{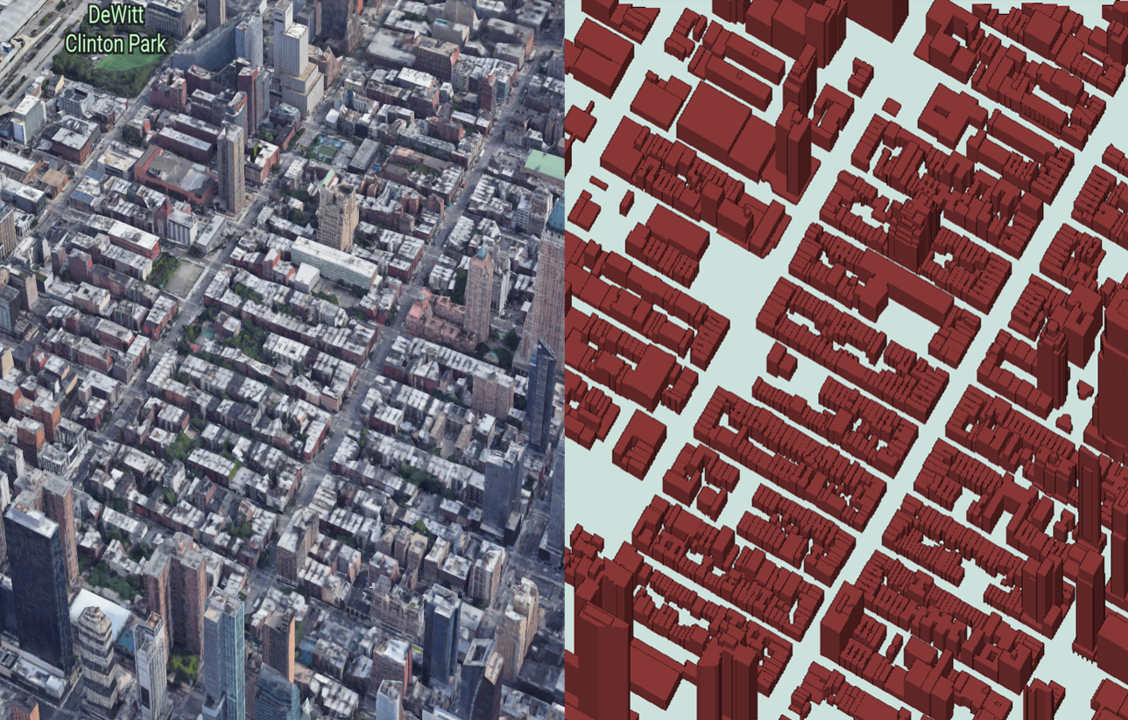}
            \caption{Aerial view of Northwest Manhattan (left) and the corresponding ray tracing model (right).}
            \label{fig:rt_NYC}
        \end{figure}

        The parameters derived from the rectangular-grid environment were validated in a more complex urban scenario of Northwest Manhattan (Fig.~\ref{fig:rt_NYC}). 
        This area features irregular building heights, spacings, and street widths, while still maintaining an overall grid-like pattern. 
        A three-dimensional map of Manhattan was converted into a CAD file using CADMAPPER~\cite{cadmapper} to run ray tracing simulations over it. 

        In this environment, APs and UEs were strategically positioned and categorized by street order (zeroth, first, and second) to compute PL values using the parameters from the rectangular-grid environment.
        Material properties and simulation parameters were kept consistent with those used in the rectangular-grid scenario. 
        This validation step was designed to assess the generalizability and robustness of the parameter estimates in a realistic urban context.

    \subsection{Measurement campaign at USC}
\begin{figure}[t!]
            \centering
            \includegraphics[width=\linewidth]{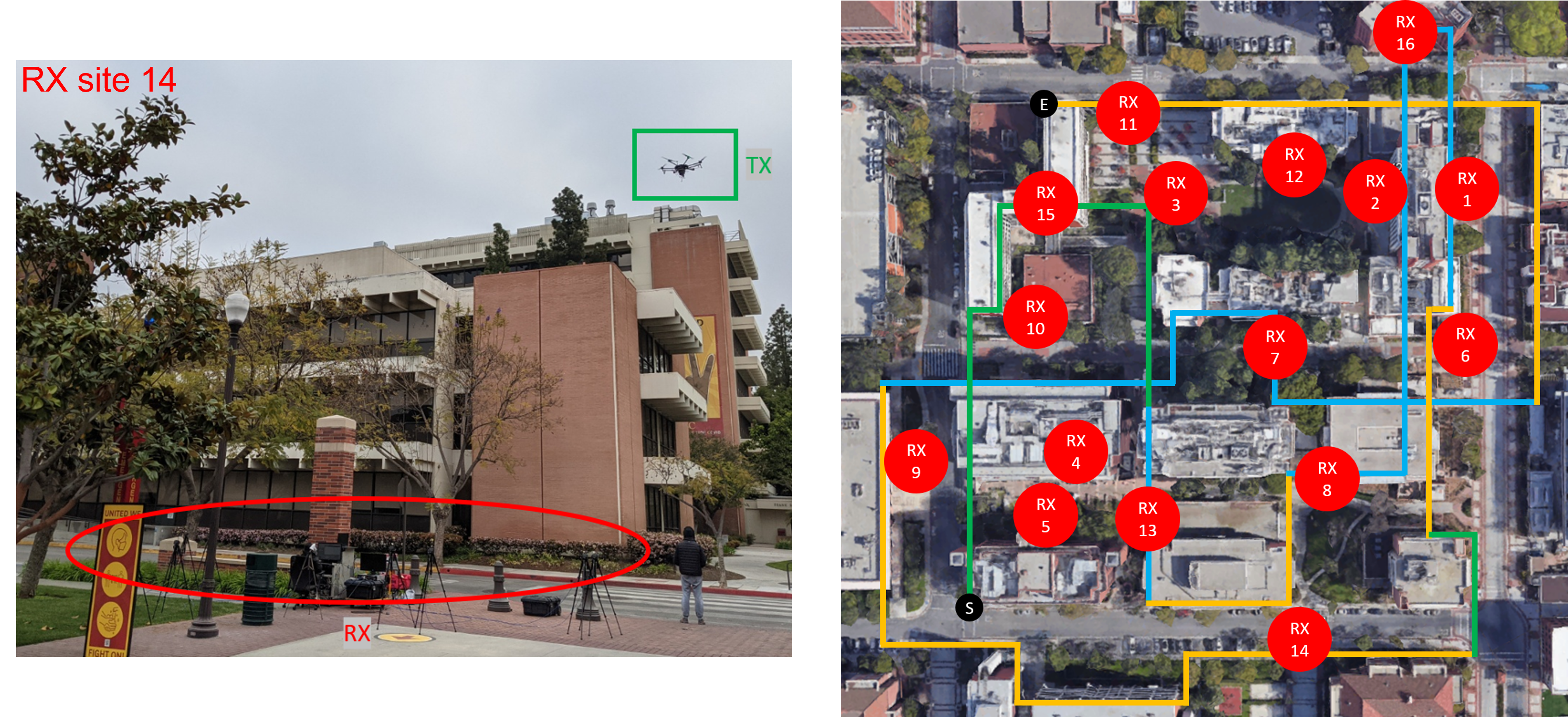}
            \caption{Left: Drone-based channel sounder (acting as AP) and ground antennas (UEs) used during the measurement campaign. 
            Right: Aerial view of the USC campus in Los Angeles, showing RX (UE) locations and TX (AP) trajectory. 
            TX trajectories at 25, 35, and 45\,m heights (yellow, green, and blue, respectively) were sampled at 5\,cm intervals, while RX antennas were placed at 1.5\,m heights.}
            \label{fig:meas}
        \end{figure}

        To capture real-world propagation effects potentially underrepresented in simulations—--such as vegetation, heterogeneous materials, and dynamic surroundings—--we conducted an outdoor CF-mMIMO channel measurement campaign on the USC University Park Campus (Fig.~\ref{fig:meas}).

        \renewcommand{\arraystretch}{1.2} 
        \begin{table}[t]
        \centering
        \caption{Summary of Measurement Setup}
        \label{tab:meas_setup}
        \begin{tabular}{p{2cm}p{5.5cm}}
        \hline
        \textbf{Component} & \textbf{Specification} \\
        \hline
        Sounding Signal & 3.5 GHz carrier frequency; 46 MHz bandwidth; 20 kHz subcarrier spacing (2301 subcarriers) \\
        TX & Drone-mounted, single omnidirectional antenna moving at 1 m/s, transmitted at 27 dBm\\
        RX & Ground-based, 8 separated omnidirectional antennas captured $8 \times 2301$ channel matrix every 50 ms ($\approx$5 cm drone step)\\
        \hline
        \end{tabular}
        \end{table}

        In this campaign, a drone-based channel sounder followed rooftop-level trajectories, sampling spatial positions every 5\,cm to emulate a dense grid of AP locations. 
        Simultaneously, eight ground-based antennas at 1.5\,m heights were deployed within a 15\,m radius at each of 16 distinct sites, collectively representing 128 UE positions (Fig.~\ref{fig:meas}, right).
        This comprehensive setup yielded PL data for over 30,000 AP locations and 128 UE positions, covering an area of \(200 \times 200\)\,m. 
        The measurement setup is summarized in Table~\ref{tab:meas_setup}, with additional information on the sounder and campaign provided in~\cite{choi2021energy, choi2022using}.

        To capture CF-mMIMO propagation characteristics at different building heights, the drone was operated at multiple altitudes during the measurements. To maintain consistency with realistic CF-mMIMO deployments, scenarios in which the drone's altitude significantly exceeded the surrounding building heights were excluded. Still, even for the lower heights of the drone, the antennas are somewhat higher than what would be used for CF-mMIMO antennas. This may lead to underestimation of the PL in particular in NLOS2 situations, with OTR contributions comparable to, or larger than, the street-guided components. 

    \begin{table*}[t]
        \centering
        \caption{Parameter Values and Statistics for the \(\alpha\)--\(\beta\) and CUNEC Models}
        \label{table:parameter}
        \begin{tabular}{c|cc|ccc|ccc}
            \toprule
            & \multicolumn{2}{c|}{\(\alpha\)--\(\beta\)} & \multicolumn{3}{c|}{CUNEC: Mean (AP / UE)} & \multicolumn{3}{c}{CUNEC: Std. Dev. (AP / UE)} \\
            \midrule
            & LOS & NLOS & Zeroth-order & First-order & Second-order & Zeroth-order & First-order & Second-order \\ 
            \midrule
            \(\Delta/\Delta_n\) & 6 & -56.2 & 6.3 & 29.7 & -- & 0.6  & 11  & -- \\
            \(\beta/\mathrm{b}_n\) & 1.58 & 6.3 & 1.56 & 1.4 & 1.3 & 0.05 & 0.65 & 0.58 \\
            \(\sigma_S/\sigma_{\mathrm{S}\text{-}N}\) & 1.2 & 11.5 & 1.1 / 0.9 & 4.8 / 4.7 & 7 / 7 & 0.11 / 0.11 & 2.6 / 1.6 & 2.4 / 1.9 \\
            \(d_{\mathrm{corr}}/\mathrm{d}_{\mathrm{corr}\text{-}N}\) & 9 & 1 & 17.8 / 12.4 & 9.9 / 14.4 & 10.4 / 16 & 6.1 / 2.8 & 6.7 / 9.5 & 10.6 / 9.2 \\
            \(\kappa\) & -- & -- & -- & 0.037 & -- & -- & 0.02 & -- \\
            \(C\) & -- & -- & -- & 9.2 & -- & -- & 4.6 & -- \\            
            \midrule
            \(d_\mathrm{th}\) & -- & -- & -- & 70 & -- & -- & -- & -- \\
            \(d/d_n\) & \(>\)15, \(<\)500 & \(>\)15, \(<\)500 & \(>\)15, \(<\)500 & \(>\)1, \(<\)250 & \(>\)1, \(<\)500 &  &  &  \\
            \bottomrule
        \end{tabular}
    \end{table*}

    \begin{table}[t]
        \centering
        \caption{Correlations of the CUNEC Model Parameters}
        \label{table:correlations}
            
        \begin{subtable}[t]{\columnwidth}
            \centering
            \caption{Zeroth-order (LOS)}
            \label{table:LOS_corr}
            \begin{tabular}{l|cccccc}
                \toprule
                & \(\mathrm{b_0}\) & \(\Delta_0\) & \(\sigma_{\mathrm{S\text{-}0,AP}}\) & \(\sigma_{\mathrm{S\text{-}0,UE}}\) & \(\mathrm{d}_{\mathrm{corr\text{-}0,AP}}\) & \(\mathrm{d}_{\mathrm{corr\text{-}0,UE}}\) \\
                \midrule
                \(b\) & -- & -- & -0.6 & -0.8 & 0.5 & 0.8 \\
                \(h\) & 0.4 & -- & -0.5 & -0.5 & 0.6 & 0.6 \\
                \(w\) & 0.9 & -0.6 & -- & -- & 0.9 & 0.8 \\
                \(\mathrm{b_0}\) & \cellcolor{gray!20} & -0.8 & -- & -- & 0.7 & 0.5 \\
                \(\Delta_0\) & \cellcolor{gray!20} & \cellcolor{gray!20} & -0.4 & -- & -0.5 & -- \\
                \(\sigma_{\mathrm{S\text{-}0,AP}}\) & \cellcolor{gray!20} & \cellcolor{gray!20} & \cellcolor{gray!20} & \cellcolor{gray!20} & -0.4 & \cellcolor{gray!20} \\
                \(\sigma_{\mathrm{S\text{-}0,UE}}\) & \cellcolor{gray!20} & \cellcolor{gray!20} & \cellcolor{gray!20} & \cellcolor{gray!20} & \cellcolor{gray!20} & -0.5 \\
                \bottomrule
            \end{tabular}
        \end{subtable}
            
        \vspace{6pt}

        \begin{subtable}[t]{\columnwidth}
            \centering
            \caption{First-order (NLOS1)}
            \label{table:NLOS1_corr}
            \begin{tabular}{l|ccccccc}
                \toprule
                & \(\mathrm{b_1}\) & \(\Delta_1\) & \(\kappa\) & \(C\) & \(\sigma_{\mathrm{S\text{-}1,AP}}\) & \(\sigma_{\mathrm{S\text{-}1,UE}}\) \\
                \midrule
                \(b\)     & -0.5 & --  & 0.6 & -- & -0.4 & -- \\
                \(h\)    & -0.5 & 0.4 & --  & --  & --   & -- \\
                \(\mathrm{b}_1\)           & \cellcolor{gray!20} & -0.9 & --  & --  & --   & -- \\
                \(\Delta_1\)      & \cellcolor{gray!20} & \cellcolor{gray!20} & --  & --  & --   & -- \\
                \(\kappa\)      & \cellcolor{gray!20} & \cellcolor{gray!20} & \cellcolor{gray!20} & 0.4 & -- & -0.4 \\
                \bottomrule
            \end{tabular}
        \end{subtable}
            
        \vspace{6pt}

        \begin{subtable}[t]{\columnwidth}
            \centering
            \caption{Second-order (NLOS2)}
            \label{table:NLOS2_corr}
            \begin{tabular}{l|ccc}
                \toprule
                & \(\mathrm{b_2}\) & \(\sigma_{\mathrm{S\text{-}2,AP}}\) & \(\sigma_{\mathrm{S\text{-}UE}}\) \\
                \midrule
                \(b\) & -0.7 & 0.5 & 0.5 \\
                \(\mathrm{b_2}\)       & \cellcolor{gray!20} & --  & -0.4 \\
                \bottomrule
            \end{tabular}
        \end{subtable}            

    \end{table}
        
\section{Path Loss Model Parameters and Statistics} \label{sec:statistics}
    This section presents the parameterization results and statistical analysis comparing the conventional deterministic \(\alpha\)--\(\beta\) model with the trajectory-specific, stochastic CUNEC approach.
    In our simulations (Sec.~\ref{sec:rect}), two distinct scenarios were explored: one involving multiple ground-level UE points with rooftop-level AP trajectories, and another involving rooftop-level AP points and ground-level UE trajectories.\footnote{Due to computational complexity, simulations involving simultaneous trajectories for both APs and UEs were not conducted, as they would require impractically long execution times.}
    For the deterministic \(\alpha\)--\(\beta\) model, \emph{all} PL data points from both scenarios are aggregated and plotted against Euclidean distance, so a single set of deterministic parameter values is retrieved.
    
    In contrast, CUNEC independently analyzes \emph{each} UE-to-AP and AP-to-UE trajectory pair to derive distributions of parameter values that inherently reflect environmental variability, preserve unique propagation characteristics, and improve the model's adaptability to complex urban scenarios.
    Moreover, CUNEC distinguishes both \(\sigma_{\mathrm{S}}\) and \(\mathrm{d}_{\mathrm{corr}}\) based on trajectory type---using \(\sigma_{\mathrm{S, AP}}\) and \(\mathrm{d}_{\mathrm{corr, AP}}\) for rooftop-level APs, and \(\sigma_{\mathrm{S, UE}}\) and \(\mathrm{d}_{\mathrm{corr, UE}}\) for ground-level UEs---to account for the distinct spatial correlation characteristics associated with proximity APs versus mobile UEs.
    
    Table~\ref{table:parameter} summarizes the key parameter values, highlighting the contrast between the deterministic nature of the \(\alpha\)--\(\beta\) model and the statistical representation inherent in CUNEC. 
    Additionally, Table~\ref{table:correlations} details the interdependencies among CUNEC model parameters, illustrating how environmental factors influence wireless propagation dynamics.
    Only correlations with a coefficient greater than 0.4 (indicating moderate correlation) are considered; parameters not shown in the table (including between parameters in the same row or column) can be regarded as uncorrelated. 

    \subsection{Zeroth-order streets (LOS)} \label{sec:zeroth}
        Zeroth-order streets represent the simplest propagation scenario, characterized by direct LOS conditions. 
        Despite this apparent simplicity, our analysis shows that the conventional \(\alpha\)--\(\beta\) model and the proposed CUNEC model capture subtle propagation variations differently. 
        Ray tracing simulations across 200 trajectories yielded approximately 50,000 PL and distance data points, providing a robust statistical basis for our comparison.
    
    
        Table~\ref{table:parameter} shows that the deterministic parameters from the \(\alpha\)--\(\beta\) model (\(\beta_\mathrm{LOS}\!=\!1.58\) and \(\Delta_\mathrm{LOS}\!=\!6\)\,dB) closely match the means obtained from CUNEC (\(\mathop{\mathbb{E}}\{\mathrm{b}_0\}\!=\!1.56\) and \(\mathop{\mathbb{E}}\{\Delta_0\}\!=\!6.3\)\,dB).
        In contrast to the $\alpha$--$\beta$ model which conflates variations {\em between} streets and {\em within} streets, CUNEC captures street-by-street variations, which make \(\mathop{\mathbb{E}}\{\sigma_{\mathrm{S}\text{-}0,\mathrm{AP}}\}\!=\!1.1\)\,dB and \(\mathop{\mathbb{E}}\{\sigma_{\mathrm{S}\text{-}0,\mathrm{UE}}\}\!=\!0.9\)\,dB slightly smaller compared to \(\sigma_{\mathrm{LOS}}\!=\!1.2\)\,dB.
        For the same reason, \(\mathop{\mathbb{E}}\{\mathrm{d}_{\mathrm{corr}\text{-}0,\mathrm{AP}}\}\!=\!17.8\)\,m and \(\mathop{\mathbb{E}}\{\mathrm{d}_{\mathrm{corr}\text{-}0,\mathrm{UE}}\}\!=\!12.4\)\,m exceed \(d_{\mathrm{corr}\text{-}\mathrm{LOS}}=9\)\,m. 
    
        Table~\ref{table:LOS_corr} reveals several correlations among the zeroth-order parameters.
        Street width (\(w\)) is positively correlated with \(\mathrm{b}_{0}\), \(\mathrm{d}_{\mathrm{corr}\text{-}0,\mathrm{AP}}\), and \(\mathrm{d}_{\mathrm{corr}\text{-}0,\mathrm{UE}}\), and negatively correlated with \(\Delta_{0}\). 
        The AP height (\(h\)) and length of a city block (\(b\)) also show meaningful correlations: they are positively correlated with both correlation distances and negatively correlated with both shadowing standard deviations, while \(h\) is additionally correlated with \(\mathrm{b}_{0}\).
        Additional correlations among model parameters are also presented, with \(\Delta_0\) exhibiting the strongest negative correlation with \(\mathrm{b_0}\).

    \subsection{First-order streets (NLOS1)}
        For first-order streets, which involve a single corner turn introducing significant diffraction and reflection losses, we analyzed 1,000 ray tracing trajectories, yielding approximately 250,000 PL and distance data points.
        The parameters derived here reflect behavior within the simulated distance range $<500$m and do not aim to reproduce a particular behavior for $d_1 \rightarrow \infty$.

        The conventional \(\alpha\)--\(\beta\) model aggregates all NLOS data---both NLOS1 and NLOS2---into a unified set of parameters (\(\Delta_{\mathrm{NLOS}}\), \(\beta_{\mathrm{NLOS}}\), \(\sigma_{S\text{-}\mathrm{NLOS}}\), and \(d_{\mathrm{corr\text{-}NLOS}}\)). 
        Table~\ref{table:parameter} presents these parameter values, highlighting a significantly steeper slope (\(\beta_{\mathrm{NLOS}}\!=\!6.3\)) compared to LOS scenarios, along with an intercept (\(\Delta_{\mathrm{NLOS}}\!=\!-56.2\)\,dB), applicable for distances beyond 15\,m. 
        The large-scale fading parameters exhibit high variability, with \(\sigma_{S\text{-}\mathrm{NLOS}}\!=\!11.5\)\,dB and \(d_{\mathrm{corr\text{-}NLOS}}\!=\!1\)\,m, highlighting rapid spatial fluctuations.

        In contrast, CUNEC explicitly distinguishes NLOS1 from NLOS2 scenarios by deriving street-order-specific parameters, including \(\Delta_1\), \(\mathrm{b}_1\), \(\sigma_{\mathrm{S\text{-}1,AP}}\), \(\sigma_{\mathrm{S\text{-}1,UE}}\) \(\mathrm{d}_{\mathrm{corr\text{-}1,AP}}\), \(\mathrm{d}_{\mathrm{corr\text{-}1,UE}}\), \(\kappa\), and \(C\) for NLOS1. 
        Table~\ref{table:parameter} summarizes the statistical properties derived from 1,000 ray tracing realizations.
        It is important to note that a direct comparison of parameters between the Euclidean-distance-based \(\alpha\)--\(\beta\) model and the Manhattan-distance-based CUNEC model may not be  appropriate, given the fundamental differences in their underlying mathematical formulations.

        Specifically, we observe an expected corner loss of \(\mathop{\mathbb{E}}\{\Delta_1\}\!=\!29.7\)\,dB and an expected propagation exponent of \(\mathop{\mathbb{E}}\{\mathrm{b}_1\}\!=\!1.4\), both exhibiting significant standard deviations of \(11\)\,dB and 0.6. 
        This variability reflects the wide range of possible PL outcomes resulting from diverse urban geometries. 
        Notably, CUNEC's expected shadowing standard deviations, \(\mathop{\mathbb{E}}\{\sigma_{\mathrm{S}\text{-}1,\mathrm{AP}}\}\!=\!4.8\)\,dB and \(\mathop{\mathbb{E}}\{\sigma_{\mathrm{S}\text{-}1,\mathrm{UE}}\}\!=\!4.7\)\,dB, are considerably lower than the \(\alpha\)--\(\beta\) model's \(\sigma_{S\text{-}\mathrm{NLOS}}\!=\!11.5\)\,dB.
        Similarly, CUNEC's average correlation distances---\(\mathop{\mathbb{E}}\{\mathrm{d}_{\mathrm{corr}\text{-}1,\mathrm{AP}}\}\!=\!9.9 \)\,m and \(\!\mathop{\mathbb{E}}\{\mathrm{d}_{\mathrm{corr}\text{-}1,\mathrm{UE}}\}\!=\!14.4\)\,m---are longer than the \(\alpha\)--\(\beta\) model's \(d_{\mathrm{corr}\text{-}\mathrm{NLOS}}\!=\!1\)\,m.
        These findings intuitively suggest that signal correlation is sustained along the street, even under NLOS conditions.
        
        Table~\ref{table:NLOS1_corr} further demonstrates how environmental parameters---such as \(b\) and \(h\)---correlate with CUNEC model parameters.
        Notably, \(\mathrm{b_1}\) decreases with increasing \(b\) and \(h\) suggesting slower increase in distance-based PL when building blocks are larger.
        Furthermore, \(b\) is negatively correlated with \(\sigma_{\mathrm{S}\text{-}1,\mathrm{AP}}\), but not correlated with \(\sigma_{\mathrm{S}\text{-}1,\mathrm{UE}}\), while \(\kappa\) shows the opposite trend - negatively correlated with \(\sigma_{\mathrm{S}\text{-}1,\mathrm{UE}}\), while not correlated with \(\sigma_{\mathrm{S}\text{-}1,\mathrm{AP}}\).
        The results underscore the spatial asymmetry between the shadowing behavior of AP and UE and reveal additional dependencies between parameters.


    \begin{figure*}[t]
        \centering
        \includegraphics[width=0.9\textwidth]{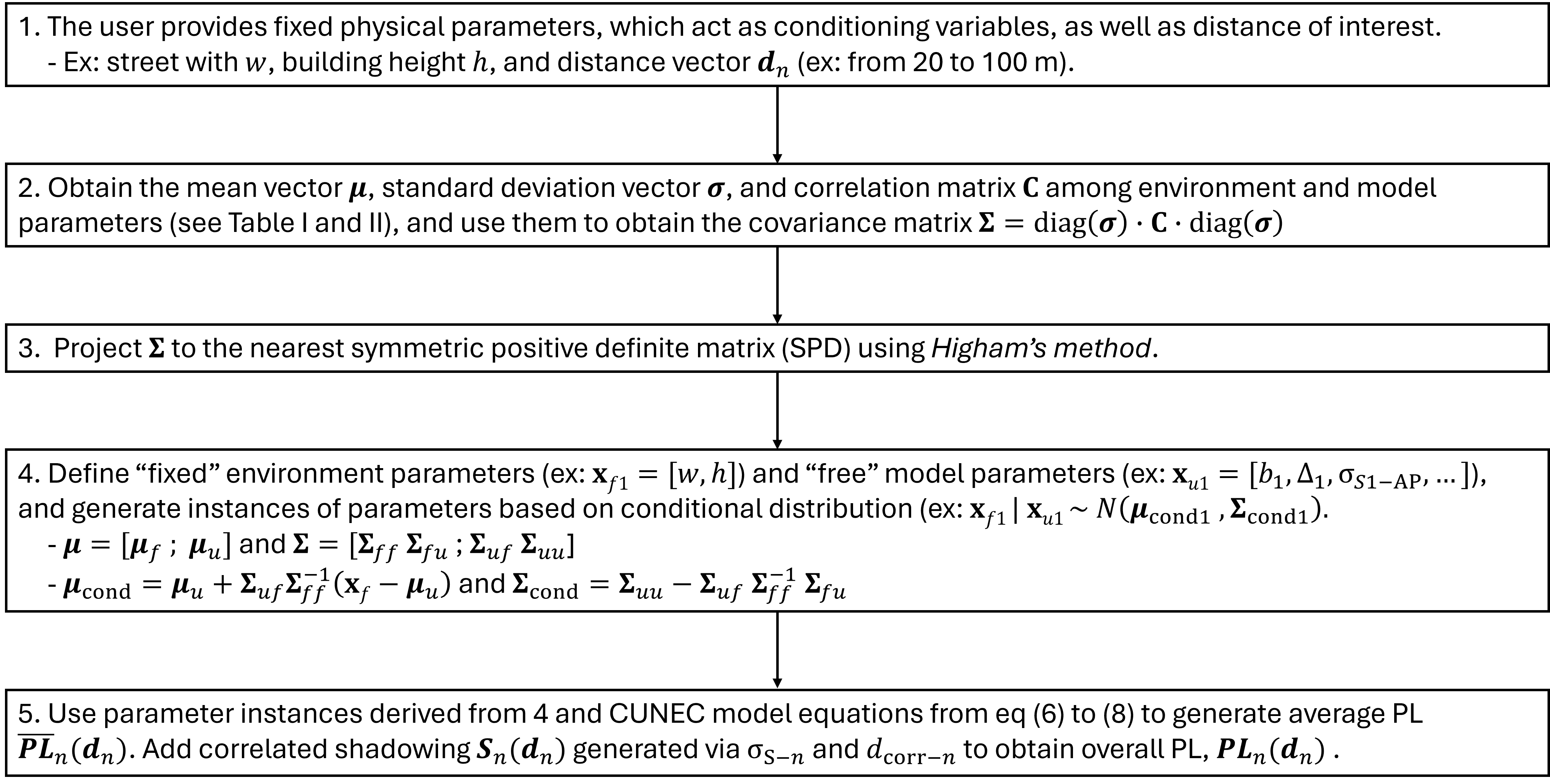} 
        \caption{Flowchart of PL instance generation for the CUNEC model.}
        \label{fig:cunec_flowchart}
    \end{figure*}

    \subsection{Second-order streets (NLOS2)}
        Second-order streets, which involve two successive corner turns, represent the most complex propagation scenarios in urban environments. 
        We analyzed 350 TX-RX trajectory pairs from ray tracing simulations, yielding approximately 90,000 PL and distance data points.
        Similar to first-order streets, the CUNEC model derives specific parameters for second-order street---\(\mathrm{b}_2\), \(\sigma_{\mathrm{S\text{-}2,AP}}\), \(\sigma_{\mathrm{S\text{-}2,UE}}\), \(\mathrm{d}_{\mathrm{corr\text{-}2,AP}}\), and \(\mathrm{d}_{\mathrm{corr\text{-}2,UE}}\). 
        Table~\ref{table:parameter} summarizes these key statistics. 
        
        The PL increases with an expected exponent of \(\mathop{\mathbb{E}}\{\mathrm{b}_2\}=1.3\).
        The high standard deviation of \(\mathrm{b}_2\), similar to NLOS1, reflects the increased variability inherent in NLOS2 propagation. 
        In addition, \(\mathop{\mathbb{E}}\{\sigma_{\mathrm{S\text{-}2,AP}}\}\!=\!\mathop{\mathbb{E}}\{\sigma_{\mathrm{S\text{-}2,UE}}\}\!=\!7\)\,dB (with standard deviations of 2.4 and 1.9\,dB) are high, though still lower than \(\sigma_{S\text{-}\mathrm{NLOS}}=11.5\)\,dB in the $\alpha$--$\beta$ model. 
        Additionally, \(\mathop{\mathbb{E}}\{\mathrm{d}_{\mathrm{corr}\text{-}2,\mathrm{AP}}\}\!=\!10.4\)\,m and \(\mathop{\mathbb{E}}\{\mathrm{d}_{\mathrm{corr}\text{-}2,\mathrm{UE}}\} \!=\!16\)\,m indicate significant spatial correlation in NLOS2 scenarios.

        Table~\ref{table:NLOS2_corr} highlights several important correlations among the NLOS2 parameters.
        Note that \(\sigma_{\mathrm{S\text{-}2,AP}}\) and \(\sigma_{\mathrm{S\text{-}2,UE}}\) correlate positively with \(b\), while correlating negatively with \(\mathrm{b}_2\).

\section{Empirical Evaluation of Models} \label{sec:emp}
    We now evaluate and compare the performance of the conventional \(\alpha\)--\(\beta\) model and the proposed CUNEC model using the parameterizations derived in Sec~\ref{sec:statistics}.
    This evaluation serves to verify both the fundamental structure of the CUNEC model and the ability of its parameterization to generalize to varying urban environments.

    We thus compare the model predictions to both the Manhattan ray tracing simulations and the USC channel sounding campaign, as described in Sec.~\ref{sec:scenarios}.
    In the CUNEC framework, model parameters are generated on a street-specific basis according to the parameterization in Sec.~\ref{sec:statistics}, conditioned on the environmental variables observed in the Manhattan and USC scenarios.
    For example, if the street width is measured at 25\,m, the corresponding statistical distributions and correlations—--both between the environmental variable and the model parameters, as well as among the model parameters themselves—--are used to generate a conditional instance of the model parameters.
    
    As illustrated in Fig.~\ref{fig:cunec_flowchart}, the procedure for generating PL instances begins by specifying fixed environmental parameters---such as $w$, $b$---which serve as conditioning inputs.
    The mean vector \( \boldsymbol{\mu} \), standard deviation vector \( \boldsymbol{\sigma} \), and correlation matrix \( \mathbf{C} \) for all model parameters, obtained from Table~\ref{table:parameter} and Table~\ref{table:correlations}, are used to compute the full covariance matrix \( \boldsymbol{\Sigma}\!=\! \text{diag}(\boldsymbol{\sigma})\!\cdot\!\mathbf{C} \!\cdot\!\text{diag}(\boldsymbol{\sigma}) \).
    This matrix is then projected to the nearest symmetric positive definite (SPD) matrix using Higham’s method~\cite{higham1988computing}.

    Based on this joint Gaussian model, parameters are partitioned into ``fixed'' environmental variables (e.g., \( \mathbf{x}_f\!=\![w, b] \)) and ``free'' model parameters (e.g., \( \mathbf{x}_u\!=\![\Delta_1, \mathrm{b}_1, \kappa, \ldots] \)). 
    Conditional multivariate Gaussian sampling is then performed to generate realizations of free parameters that preserve the conditional statistics and correlations associated with the given environment.
    To ensure physical plausibility, sampled values are constrained by applying lower bounds to parameters that must remain positive—such as PL exponent, shadowing standard deviations, correlation distances, etc.
    These sampled parameters are substituted into the CUNEC model equations (e.g., \eqref{eq:PL0} to \eqref{eq:PL2}) to compute the average component of PL, which is combined with generated spatially correlated shadowing to yield the final predicted PL realizations.
    In contrast, the \(\alpha\)--\(\beta\) model employs fixed parameter values for each scenario, with all variations in PL attributed solely to large-scale fading, \(\sigma_S\). 

    \subsection{Northwest Manhattan}
        \begin{figure*}[t]
          \centering
          \begin{minipage}[t]{0.4\textwidth}
            \centering
            \subcaptionbox{LOS: AP trajectory\label{subfig:AP_LOS}}[\textwidth]{%
              \includegraphics[width=\textwidth]{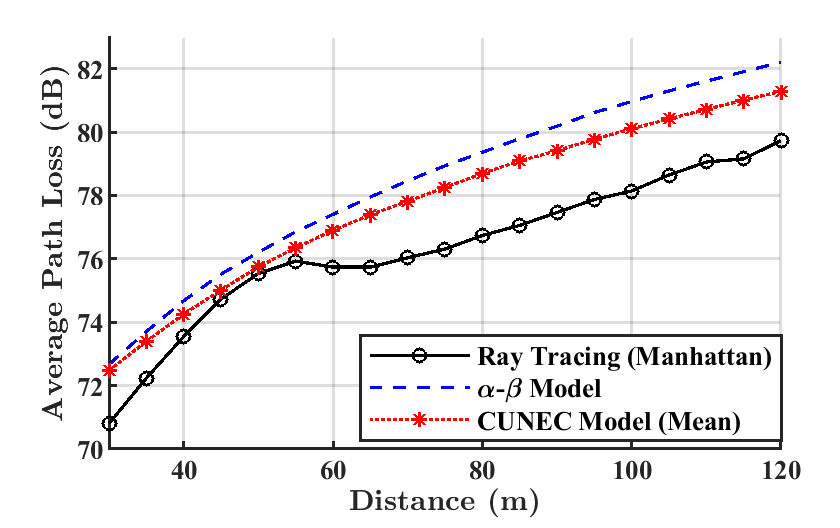}%
            } 
            \subcaptionbox{NLOS1: AP trajectory\label{subfig:AP_NLOS1}}[\textwidth]{%
              \includegraphics[width=\textwidth]{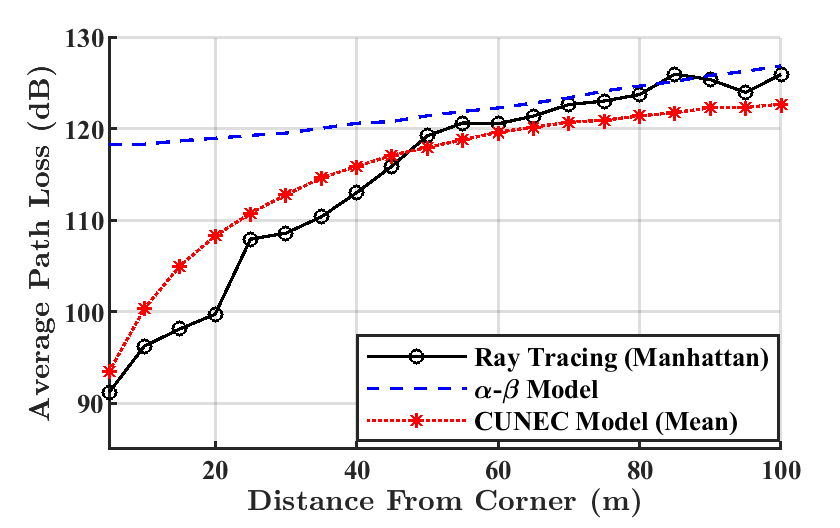}%
            }
            \subcaptionbox{NLOS2: AP trajectory\label{subfig:AP_NLOS2}}[\textwidth]{%
              \includegraphics[width=\textwidth]{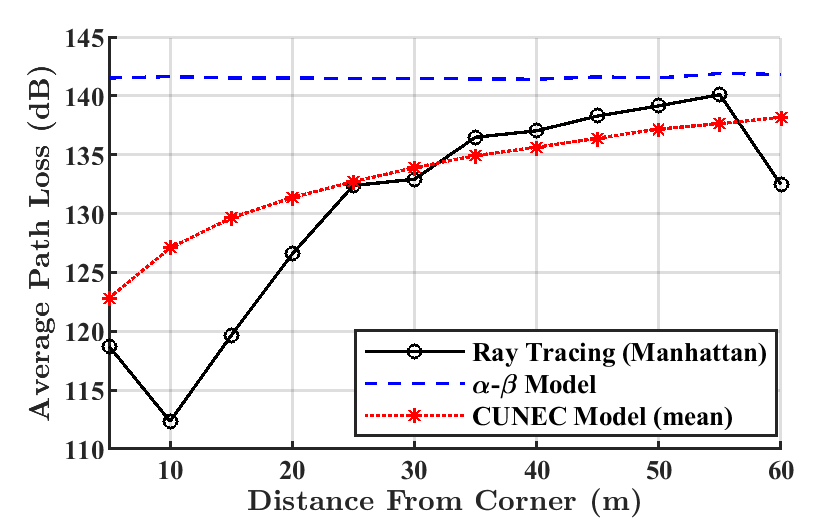}%
            }
          \end{minipage}
          \hfill
          \begin{minipage}[t]{0.4\textwidth}
            \centering
            \subcaptionbox{LOS: UE trajectory\label{subfig:UE_LOS}}[\textwidth]{%
              \includegraphics[width=\textwidth]{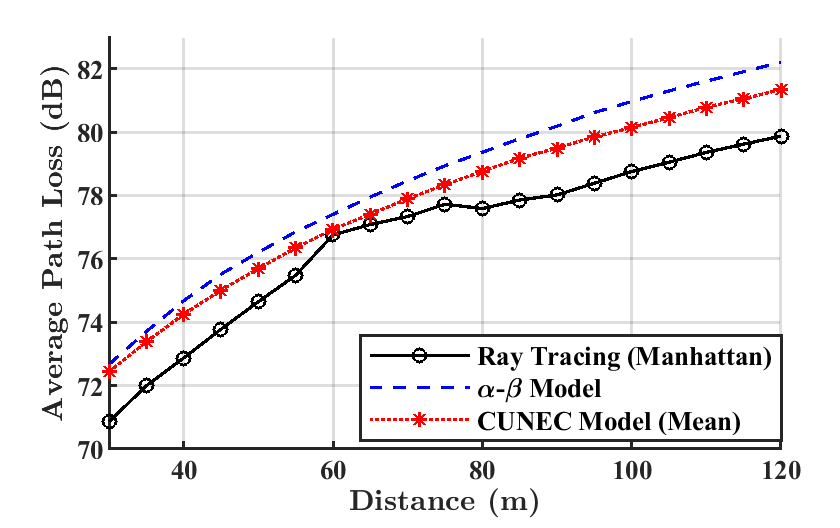}%
            } \label{fig:UE_LOS}
            \subcaptionbox{NLOS1: UE trajectory\label{subfig:UE_NLOS1}}[\textwidth]{%
              \includegraphics[width=\textwidth]{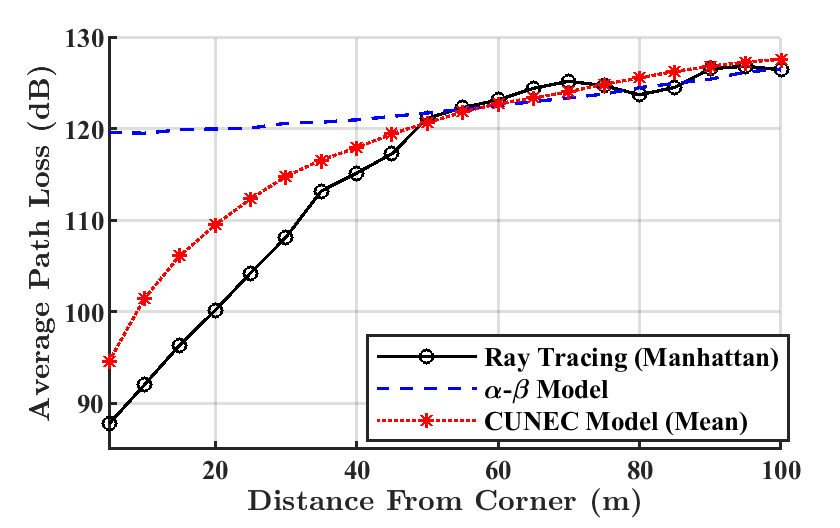}%
            }
            \subcaptionbox{NLOS2: UE trajectory\label{subfig:UE_NLOS2}}[\textwidth]{%
              \includegraphics[width=\textwidth]{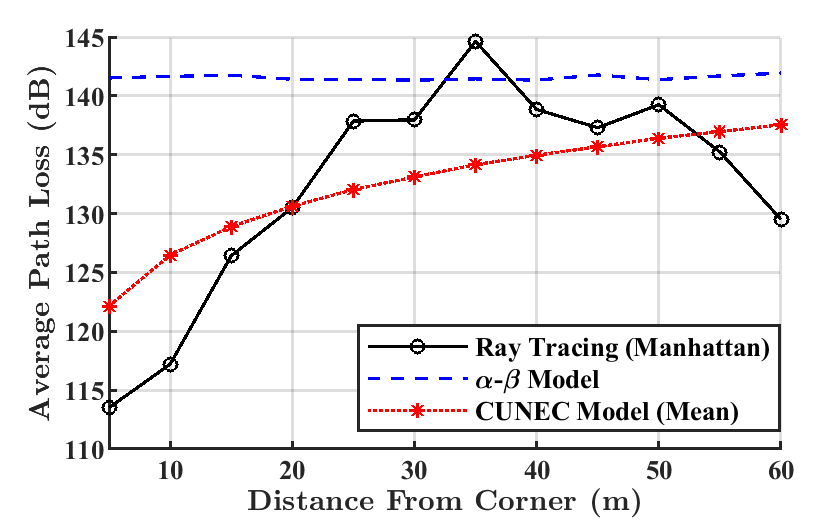}%
            }
          \end{minipage}
          \caption{
            PL versus distance for LOS, NLOS1, and NLOS2, comparing the \(\alpha\)--\(\beta\) and CUNEC models against ray tracing simulations in a Manhattan environment.
            The left column shows the AP trajectory; the right column shows the UE trajectory.
          }
          \label{fig:AP_UE}
        \end{figure*}

        Fig.~\ref{fig:AP_UE} compares the mean PL as a function of distance from the Manhattan ray tracing simulations with the corresponding predictions from the \(\alpha\)--\(\beta\) and CUNEC models.
        For the Manhattan scenario, we consolidated data from thirty fixed UE-to-AP trajectory combinations and thirty fixed AP-to-UE trajectory combinations. 
        These trajectories were then averaged to produce the diamond markers shown in Fig.~\ref{fig:AP_UE}. 
        
        We first note that the mean PL predictions from the \(\alpha\)--\(\beta\) and CUNEC models are closely aligned for LOS cases (for streets with \(w\!=\!20\)\,m and \(h\!=\!10\)\,m) in Fig.~\ref{subfig:AP_LOS} and Fig.~\ref{subfig:UE_LOS}, reflecting their similar parameter values and shared reliance on Euclidean distance in the LOS regime. 
        However, while subtle, we notice the CUNEC model matches better with the ray tracing than the \(\alpha\)--\(\beta\) model, as CUNEC depends on environment variables like street width, which is correlated with the PL exponent, as shown in Table~\ref{table:LOS_corr}.
        The root mean squared error (RMSE) between the  \(\alpha\)--\(\beta\) model and ray tracing was 2.2 dB for both the AP and UE trajectories, whereas CUNEC achieved substantially lower values of 1.5 dB and 1.6 dB. Notably, under the best-case instance, the MSE with CUNEC further decreased to 0.9 dB.
        
        For NLOS1, Figs.~\ref{subfig:AP_NLOS1} and \ref{subfig:UE_NLOS1} show the mean PL as a function of distance from the corner, for streets with \(w\!=\!20\)\,m and \(d_{\mathrm{c}}\!=\!80\)\,m. 
        Each plot aggregates data from ten distinct AP--UE trajectory pairs.
        Notably, the \(\alpha\)--\(\beta\) model shows only a slight increase in PL with distance, because the Euclidean distance between the AP and UE grows slowly despite the growing Manhattan distance. 
        Despite its higher NLOS PL exponent, the \(\alpha\)--\(\beta\) model cannot capture the steep rise observed in the ray tracing simulations, where the mean PL grows from approximately 90\,dB to 125\,dB across 100\,m range. In contrast, the CUNEC model captures this increase.
        In NLOS1, the RMSE between the \(\alpha\)--\(\beta\) model and ray tracing was 12 dB for both the AP and UE trajectories. In contrast, CUNEC yielded markedly lower errors of 2.2 dB, which further decreased to 1.7 dB and 1.8 dB in the best-case instance.
        
        Figs.~\ref{subfig:AP_NLOS2} and \ref{subfig:UE_NLOS2} illustrate the second-order (NLOS2) scenario, with \(w\!=\!20\)\,m, distance from TX to first corner of 20\,m, and distance from first corner to second corner of 210\,m. 
        We used 12 unique trajectory instances for both the AP and UE plots. 
        Here, the PL observed in the measured data does not follow a strictly monotonic trend.
        Multiple corner diffractions and partial waveguiding effects introduce additional reflection paths, making the overall PL become more sensitive to specific distances and angles around each corner. 
        While the ray tracing simulations exhibit these nuances, the \(\alpha\)--\(\beta\) model, similar to NLOS1, relies on a single Euclidean distance-based exponent and thus fails to capture the more complex progression of PL, exhibiting near constant PL behavior. 
        In contrast, CUNEC’s second-order street classification and the dedicated PL exponent help better represent the rising PL curve, especially near the corner.
        For the NLOS2 scenario, the \(\alpha\)--\(\beta\) exhibited large RMSE of 13 dB for the AP and UE trajectories. CUNEC, however, substantially mitigated the error, attaining 5.5 dB and 6.5 dB on average, and as low as 3.7 dB and 5.2 dB in the most favorable case.
        

    \subsection{Measurement campaign at USC}
        \begin{figure}[t]
            \centering
            \subcaptionbox{LOS: AP trajectory\label{subfig:meas_LOS}}[0.8\linewidth]{%
                \includegraphics[width=\linewidth]{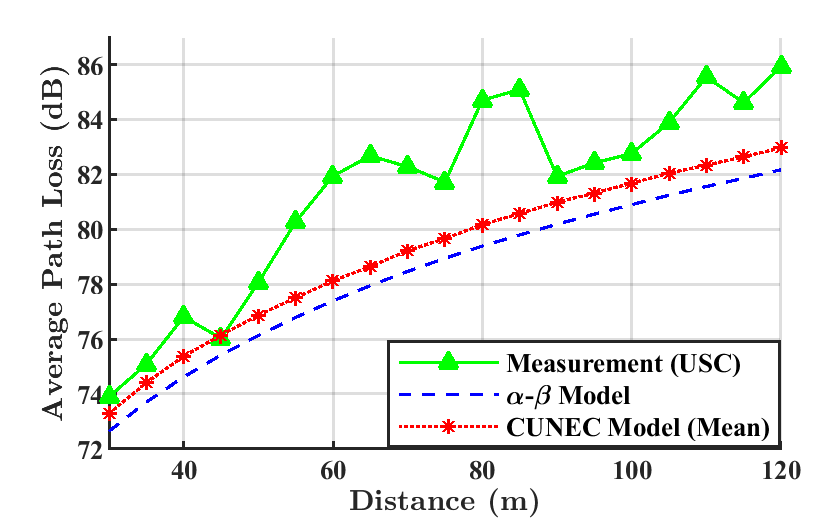}%
            }
            \subcaptionbox{NLOS1: AP trajectory\label{subfig:meas_NLOS1}}[0.8\linewidth]{%
                \includegraphics[width=\linewidth]{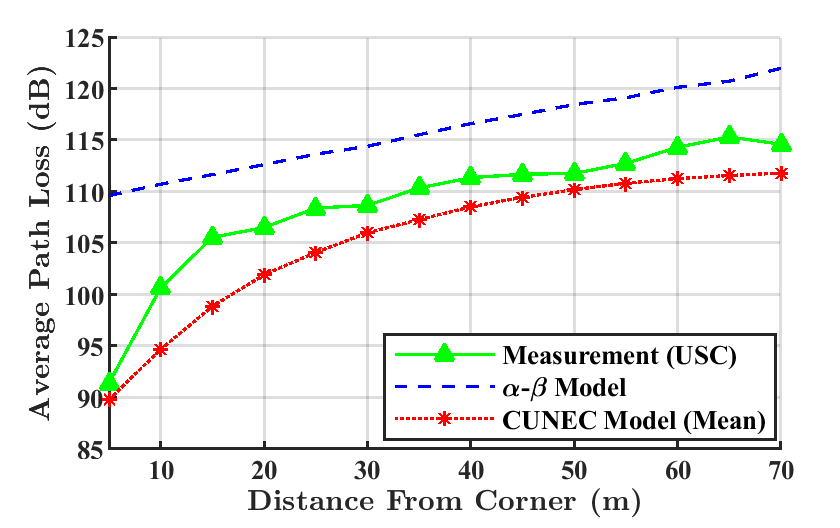}%
            }
            \subcaptionbox{NLOS2: AP trajectory\label{subfig:meas_NLOS2}}[0.8\linewidth]{%
                \includegraphics[width=\linewidth]{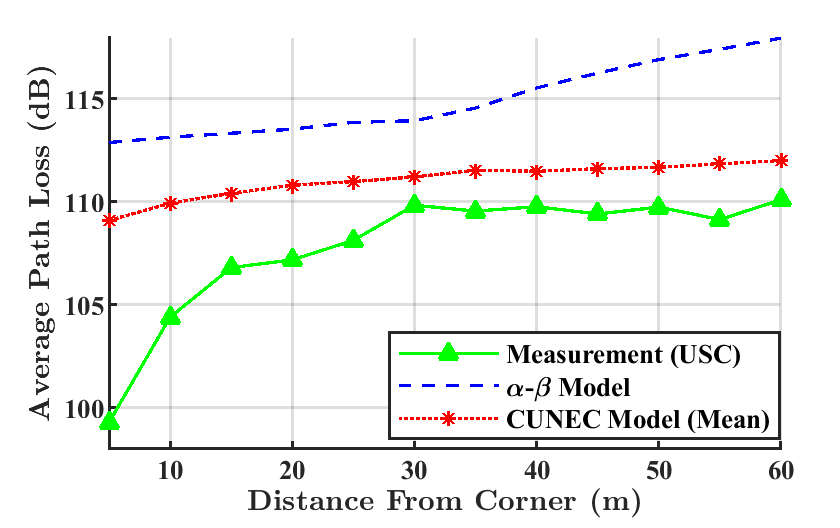}%
            }
            \caption{PL versus distance for LOS, NLOS1, and NLOS2, comparing the \(\alpha\)--\(\beta\) and CUNEC models against channel measurement data in a USC setting.}
            \label{fig:meas_scenarios}
        \end{figure}
        
        Fig.~\ref{fig:meas_scenarios} presents PL measurements from a channel measurement campaign conducted on the USC campus, alongside predictions from both the \(\alpha\)--\(\beta\) and CUNEC models. 
        As before, we evaluate three scenarios---LOS, NLOS1, and NLOS2---each plotted as a function of the distance from the corner along an AP trajectory.
        The UE trajectory is omitted due to the measurement methodology discussed in Sec. IV.C.
        
        In Fig.~\ref{subfig:meas_LOS}, we selected eight measurement instances with \(w=25\)\,m.
        Because of the difficulty of maintaining a clear LOS, even the cases where several trees introduced obstruction are labeled as LOS.
        The LOS PL increases less regularly with distance, and the error in the estimated PL is larger (up to 6\,dB); however, similar to the Manhattan ray tracing results, both the \(\alpha\)--\(\beta\) and CUNEC models capture the overall trend.
        In the LOS scenario, the \(\alpha\)--\(\beta\) model yielded a RMSE of 3.3 dB, whereas CUNEC reduced the error to 2.6 dB on average and further to 1.5 dB in the most favorable instance.
               
        Fig.~\ref{subfig:meas_NLOS1} shows the NLOS1 measurement results, where a single corner turn introduces moderate diffraction losses. 
        Eight instances with \(w=10\)\,m and \(d_\mathrm{c}=70\)\,m were chosen.
        Again, the \(\alpha\)--\(\beta\) model, which is solely based on Euclidean distance, overestimates the mean PL. 
        In contrast, CUNEC aligns much better with the measured data, reflecting its explicit handling of corner diffraction and street-order classification.  
        In the NLOS1 scenario, the \(\alpha\)--\(\beta\) model produced a RMSE of 6.8 dB, while CUNEC lowered the error to 4.1 dB on average and to 2.2 dB in the best-case instance.
        
        In Fig.~\ref{subfig:meas_NLOS2}, the NLOS2 measurements reveal a rise in PL, consistent with the expected effects of multiple diffractions and reflections around two corners. 
        Eight instances with \(w=20\)\,m,  distance from TX to first corner of 30\,m, and distance from first corner to second corner of 80\,m were chosen.
        Note that the measured PL may appear lower in some cases because the drone occasionally flew above rooftop height,  allowing ORT propagation that reduced the overall PL---an effect not captured by the CUNEC model in its current form. Still, CUNEC tracks the measured data more closely.
        In the NLOS2 scenario, the \(\alpha\)--\(\beta\) resulted in a MSE of 7.4 dB, whereas CUNEC reduced the error to 4.8 dB on average and further to 1.7 dB in the best-case instance.
        

        \begin{table}[t]
            \centering
            \caption{Local Parameter Statistics [mean (Std. Dev.)]}
            \label{tab:all_scenarios}
        
            \begin{subtable}[t]{\linewidth}
                \centering
                \caption{LOS}
                \label{tab:los_stats}
                \begin{tabular}{lcccc}
                    \toprule
                    \textbf{Method} & \(\alpha\) & \(\beta\) & \(\sigma_{S}\) & \(d_{\mathrm{corr}}\) \\
                    \midrule
                    CUNEC   & 49.4 (0.9) & 1.58 (0.06) & 0.38 (0.08) & 7.36 (2.7) \\
                    Manhattan  & 53.5 (1.6) & 1.25 (0.1) & 1.2 (0.8) & 7.75 (1.9) \\
                    USC  & 48.4 (3.9) & 1.71 (0.24) & 2.9 (0.7) & 8 (1.4) \\
                    \bottomrule
                \end{tabular}
            \end{subtable}
        
            \vspace{2mm}
        
            \begin{subtable}[t]{\linewidth}
                \centering
                \caption{NLOS1 - Manhattan}
                \label{tab:nlos1_stats}
                \begin{tabular}{lccc}
                    \toprule
                    \textbf{Method} & \(\beta\) & \(\sigma_{S}\) & \(d_{\mathrm{corr}}\) \\
                    \midrule
                    CUNEC   & 5.4 (1.1) & 7.5 (3.1) & 14.5 (8.8) \\
                    Manhattan & 5.4 (0.12) & 6.2 (1.2) & 25.3 (8.6) \\
                    \bottomrule
                \end{tabular}
            \end{subtable}
        
            \vspace{2mm}

            \begin{subtable}[t]{\linewidth}
                \centering
                \caption{NLOS1 - USC}
                \label{tab:nlos1_stats_meas}
                \begin{tabular}{lccc}
                    \toprule
                    \textbf{Method} & \(\beta\) & \(\sigma_{S}\) & \(d_{\mathrm{corr}}\) \\
                    \midrule
                    CUNEC   & 5.3 (0.92) & 7.5 (3.7) & 8.4 (3.4) \\
                    USC  & 5.4 (0.03) & 4.2 (1.1) & 5 (1.2) \\
                    \bottomrule
                \end{tabular}
            \end{subtable}
        
            \vspace{2mm}
        
            \begin{subtable}[t]{\linewidth}
                \centering
                \caption{NLOS2 - Manhattan}
                \label{tab:nlos2_stats}
                \begin{tabular}{lccc}
                    \toprule
                    \textbf{Method} & \(\beta\) & \(\sigma_{S}\) & \(d_{\mathrm{corr}}\) \\
                    \midrule
                    CUNEC   & 5.1 (1.1) & 8.6 (2.2) & 7.1 (3.1) \\
                    Manhattan  & 5.4 (0.26) & 11 (3.5) & 8.3 (3.4) \\
                    \bottomrule
                \end{tabular}
            \end{subtable}

            \vspace{2mm}

            \begin{subtable}[t]{\linewidth}
                \centering
                \caption{NLOS2 - USC}
                \label{tab:nlos2_stats_meas}
                \begin{tabular}{lcccc}
                    \toprule
                    \textbf{Method} & \(\beta\) & \(\sigma_{S}\) & \(d_{\mathrm{corr}}\) \\
                    \midrule
                    CUNEC   & 5 (1.1) & 9.2 (2.9) & 16.6 (9) \\
                    USC  & 4.9 (0.05) &  2.4 (0.3) & 26 (3) \\
                    \bottomrule
                \end{tabular}
            \end{subtable}
        \end{table}

    \subsection{Local Parameter Statistics}
    \label{sec:local_stats}
    
    While the global (i.e., averaged) PL results indicate that CUNEC aligns well with both the Manhattan ray tracing and the USC measurement, such agreement does not necessarily ensure that individual realizations (or local statistics) match those of the real-world environments. 
    To investigate this further, we fit a linear function to each realization’s PL versus \(\log_{10}\) of the \emph{Manhattan distance} for LOS, NLOS1, and NLOS2.
    In other words, we fit $\alpha$--$\beta$ model to each instance (because $\alpha$ varies highly in NLOS cases, we fix $\alpha\!=\!0$ for NLOS so the error deviation coming from $\beta$ can be observed more easily). We stress that this linear fit is used only to {\em evaluate} the models; it is not used to {\em be} another model. Furthermore, the question we wish to answer is only whether the such-obtained fits take on similar parameter values in the parameterized CUNEC model, Manhattan results, and USC; this allows us to analyze the {\em robustness} of the parameter estimates. It is {\em not} intended to evaluate the quality of the fit in each case. 
    
    Table~\ref{tab:all_scenarios} summarizes the resulting local parameter statistics. 
    In particular, Table~\ref{tab:los_stats} focuses on LOS. Since the environmental parameters (Street width etc.) are similar for Manhattan and USC, the CUNEC model takes on approximately the same parameter values.
    The results show that CUNEC, ray tracing, and measurement data exhibit similar parameter values, confirming that even for the statistics of the per-realization fits, the LOS model in CUNEC remains consistent with the measured scenarios.
    
    For NLOS1, we differentiate between the Manhattan and USC environments because the relevant environmental parameters differed. 
    Table~\ref{tab:nlos1_stats} and \ref{tab:nlos1_stats_meas} reveal that the PL exponent \(\beta\) are very similar. 
    The maximum deviation in PL, when using mean parameters as a function of distance, is no more than 3.3\,dB.
    In addition, CUNEC exhibits a slightly higher shadowing standard deviation. 
    We also note that \(d_{\mathrm{corr}}\) is higher in Manhattan, albeit with a larger standard deviation.
    Meanwhile, the USC measurement data show a lower \(d_{\mathrm{corr}}\). 
    
    
    For NLOS2, Table~\ref{tab:nlos2_stats} and \ref{tab:nlos2_stats_meas} indicate that for the cases considered, the PL exponent mean reduces slightly, but the shadowing standard deviation increases significantly.
    Similar to NLOS1, the model can follow the trend of mean correlation distance increasing in the USC scenario while decreasing in Manhattan scenario.
    Despite the discrepancies in the shadowing standard deviation statistics, the mean PL values remain within a few dB of the measured data. 
    
    %

\section{Limitations}

While the CUNEC model provides detailed modeling of CF-mMIMO channels, it has a number of limitations that must be kept in mind when using it in simulations:
\begin{itemize}[leftmargin =*]
    \item The model is tailored for environments with regular, Manhattan-like street grids. It does not explicitly account for irregular features such as oblique street intersections or open plazas, which are common in many European cities. As a result, its accuracy may degrade in those settings where the grid structure is less regular.
    \item The parameterization is done for a frequency of 3.5 GHz, as this is most likely to be used for CF-mMIMO systems. However, model parameters might be different at different operating frequencies, i.e., pathloss might change beyond the usual $f^2$ dependence.
    \item Parameters are fitted for an operating range $\le 500$m. The model should not be applied outside that range of validity.
    \item The modeling of PL when turning a corner is based on observations, and does not aim to reproduce the exact behavior arising from the combination of diffraction and reflections \cite{erceg1994diffraction}. 
    \item The models are based on propagation between isotropic antennas. Use of directional antennas would clearly change the PL behavior. 
\end{itemize}

Some of these limitations will be tackled in our future work. We stress that despite these limitations, {\em  CUNEC is more general and realistic than any state-of-the-art model}, incorporating for the first time the most critical peculiarities of CF-mMIMO channels in urban microcells. 
       

\section{Potential applications}

While this paper has focused on the details of the model and its implementation, we here briefly outline potential applications for system design (since system design itself is beyond the scope of this paper, we do not provide a detailed literature review of this topic, but refer to the extensive survey \cite{demir2021foundations} and references therein). Firstly, the model can be applied for obtaining more realistic assessments of the system capacity as a function of AP density. While capacity as function of AP density has been explored in a number of papers, but under the assumption of Euclidean PL law, as well as two-dimensional PPP, i.e., locations were not restricted to the street canyons, as is common in urban scenarios. With our model, both regular placement or random placement of one or both of APs and UEs can be analyzed. A related quantity of interest is the capacity as function of the cluster size, where thanks to our model it becomes possible to analyze, e.g., clustering of APs that are in the same street versus clusters that encompass APs in intersecting streets. In all these circumstances, the performance metric (capacity) depends on the signal and the interference power, which all can be obtained from our model, with just the necessity of ``labeling" the contributions correctly. Further applications include the analysis of energy efficiency of different arrangements \cite{choi2021energy}. 

\section{Data Release} \label{sec:data_release}
    Channel data from the USC measurements are publicly available through \cite{choi2025usc}.
    This dataset provides comprehensive channel measurement results from urban propagation environments, including LOS, NLOS1, and NLOS2 scenarios.
    Besides serving to validate the results presented in this paper, it can also support diverse research applications, such as validating and refining propagation models, applying machine learning techniques to predict channel behavior and optimize communication systems, analyzing CF-mMIMO deployments for antenna placement and spatial correlation effects, and investigating the influence of environmental features on channel characteristics.
    
    The dataset includes: 
    \begin{itemize} [leftmargin=*]
        \item \textbf{PL data}: PL data for 30,000 TX to 128 RX pairs within a 200\,m \(\times\) 200\,m area, centered at 3.5~GHz with 40~MHz bandwidth, calibrated to eliminate sounder effects. 
        \item \textbf{GPS data}: GPS coordinates of TX and RX during measurements, enabling mapping on platforms like Google Maps. This allows for detailed analysis of MPC interactions and their correlation with PL values.
    \end{itemize}

    Furthermore, MATLAB code for generating channel realizations based on CUNEC model is available via \cite{choi2025cunec}.
    The code will be based on parameterization shown in Table~\ref{table:parameter} and \ref{table:correlations}, but can be modified by the users based on their specific requirements.
    
\section{Conclusion} \label{sec:conclusion}
    This work introduced CUNEC, a novel PL model designed to significantly improve the accuracy of real-world channel characterization for CF-mMIMO systems.
    By explicitly incorporating environment-specific propagation characteristics across LOS, NLOS1, and NLOS2 conditions, CUNEC addresses the limitations of the traditional $\alpha$--$\beta$ model, offering a more precise representation of how both environmental factors and AP/UE geometry influence PL.
    
    Our validation, using ray tracing simulations in Manhattan and real-world channel measurements at USC, demonstrates CUNEC's superior accuracy in modeling signal attenuation and propagation dynamics.
    To further its impact, we have open-sourced the channel data, as well as a program for generating CUNEC-based channel data.
    
    Beyond improving PL modeling, CUNEC can serve as a practical reference for network engineers, system designers, and researchers involved in future system simulations and experimental studies of CF-mMIMO networks and bridge the gap between investigations with simple Euclidean models and analysis tied to a particular location and deployment. 

\bibliography{IEEEabrv,references.bib}
\bibliographystyle{IEEEtran}

\begin{IEEEbiography}[{\includegraphics[width=1in,height=1.25in,clip,keepaspectratio]{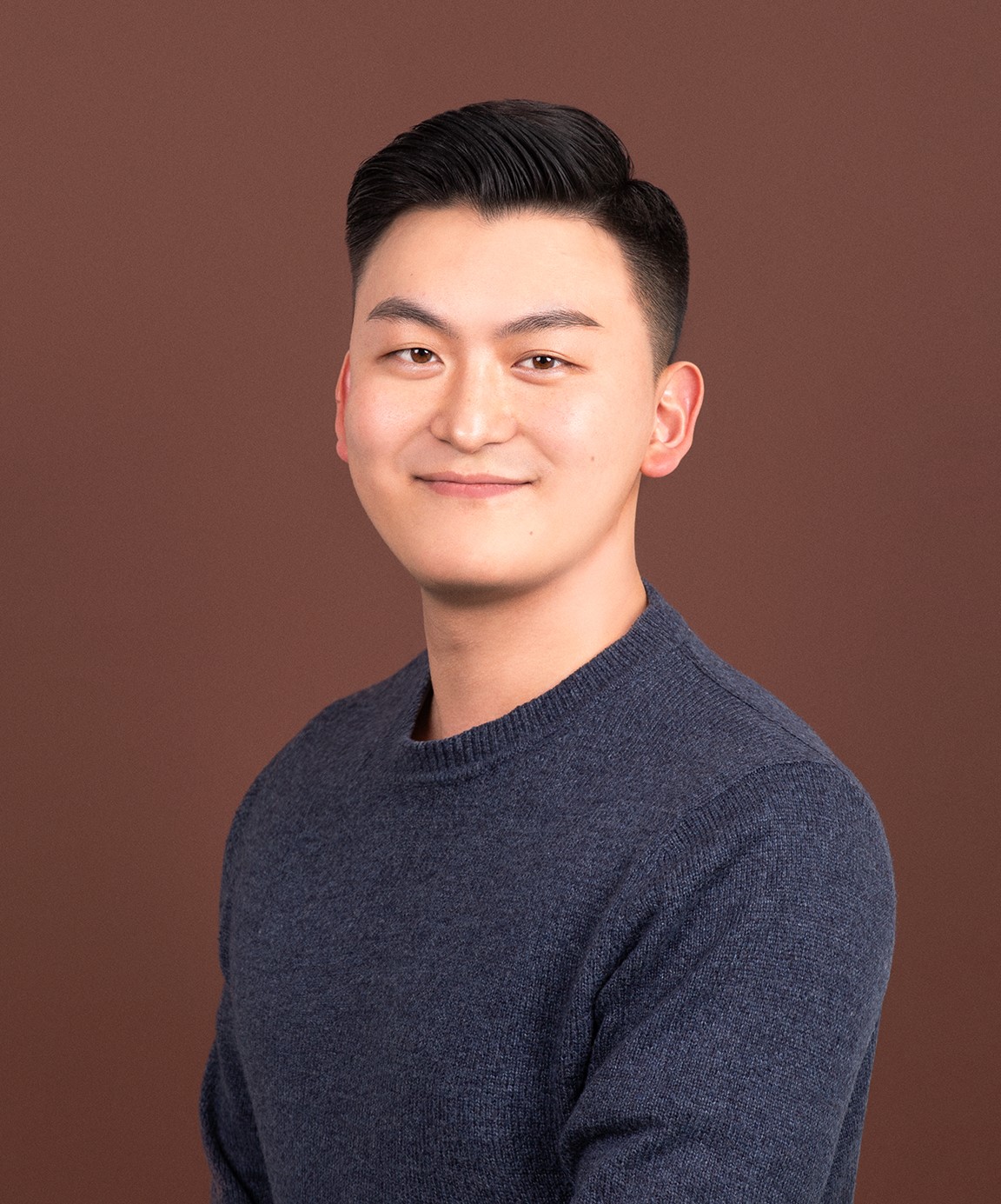}}]{Thomas Choi} received the Ph.D. degree in Electrical Engineering from the University of Southern California in 2025, where his research centered on drone-based and large-scale massive MIMO channel measurements, propagation modeling, and system-level wireless analysis. From 2022 to 2025, he was a Cellular Systems Engineer at Google, working on modem power modeling and performance characterization across real-world operating conditions. He is currently a Senior Wireless Systems Engineer at Apple, where he focuses on data-driven user-experience modeling, wireless-system validation, and measurement-to-simulation correlation for next-generation audio and connectivity systems. His research interests include channel modeling, user-centric communications, wireless system optimization, and machine-learning-based analysis of RF performance.
\end{IEEEbiography}

\begin{IEEEbiography}[{\includegraphics[width=1in,height=1.25in,clip,keepaspectratio]{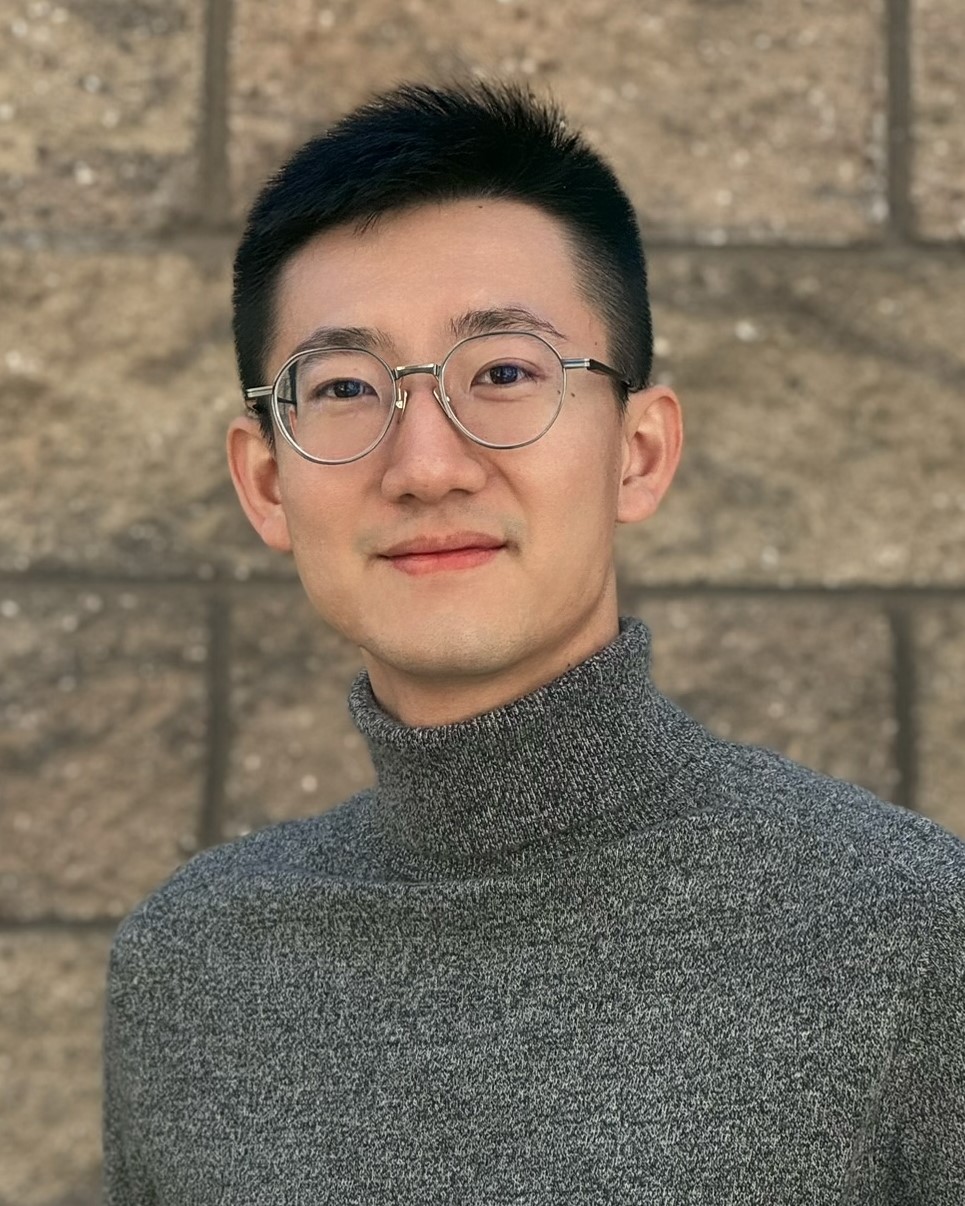}}]{Yuning Zhang} received the B.S. degree in Electrical Information Engineering from Beihang University, Beijing, China, in 2018 and the M.S. degree in Electrical Engineering from the University of Southern California, Los Angeles, CA, USA, in 2020. From 2020 to 2021, he was a resource employee for the University of Southern California (USC) Wireless Devices and Systems (WiDeS) group. He is currently pursuing a Ph.D. degree in Electrical Engineering at USC. His research interests include vehicle-to-vehicle (V2V) communication, distributed MIMO systems, localization, and mmWave channel measurement-based modeling and analysis. He also holds a patent.
\end{IEEEbiography}

\begin{IEEEbiography}[{\includegraphics[width=1in,height=1.25in,clip,keepaspectratio]{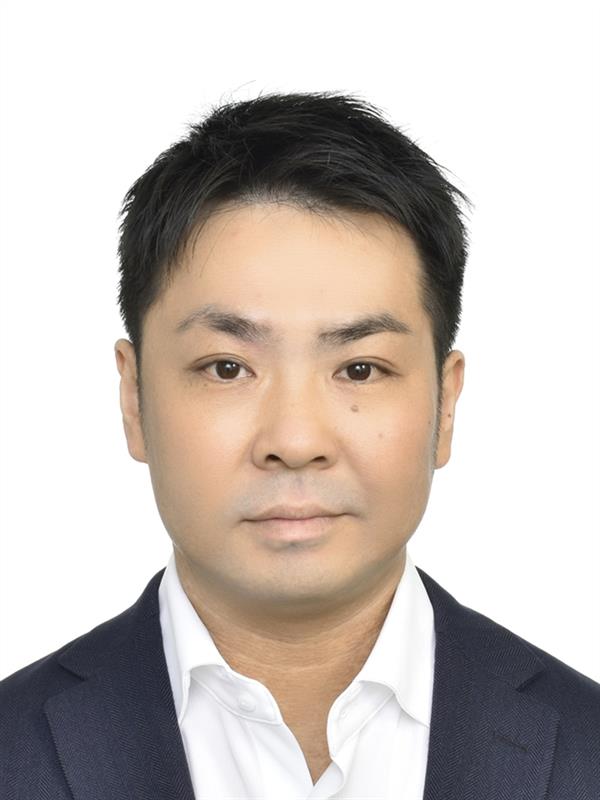}}]{Issei Kanno} received the Ph.D. degree from the Tokyo Institute of Technology, Tokyo, Japan, in 2008. He then joined KDDI Corporation, where he has been engaged in research on signal processing, antennas, and propagation in mobile communication systems. From 2013 to 2015, he was engaged in research on spectrum sharing technologies at the Advanced Telecommunications Research Institute International (ATR). Since 2015, he has been engaged in research on wireless communication systems, including 5G and 6G, at KDDI Research, Inc. Since 2021, he has been a Senior Manager of the Wireless Communications System Laboratory at KDDI Research, Inc. He has also been contributing as a subleader of the 6G Radio Technology Project in the XG Mobile Promotion Forum (XGMF) since 2024.
\end{IEEEbiography}

\begin{IEEEbiography}[{\includegraphics[width=1in,height=1.25in,clip,keepaspectratio]{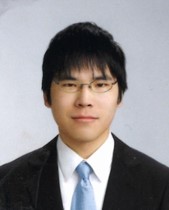}}]{Masaaki Ito} received the B.S. and M.S. degrees in wireless communications from Waseda University, Tokyo, Japan, in 2016 and 2018, respectively. From 2018 to 2019, he was with KDDI Corporation, Tokyo, Japan, where he was engaged in mobile network operations. From 2021 to 2023, he was a Visiting Researcher with the University of Southern California, Los Angeles, CA, USA. He is currently a Core Researcher with the Wireless Communications System Laboratory, KDDI Research Inc., Saitama, Japan. His research interests include user-centric networks and millimeter-wave wireless communication systems.
\end{IEEEbiography}

\begin{IEEEbiography}[{\includegraphics[width=1in,height=1.25in,clip,keepaspectratio]{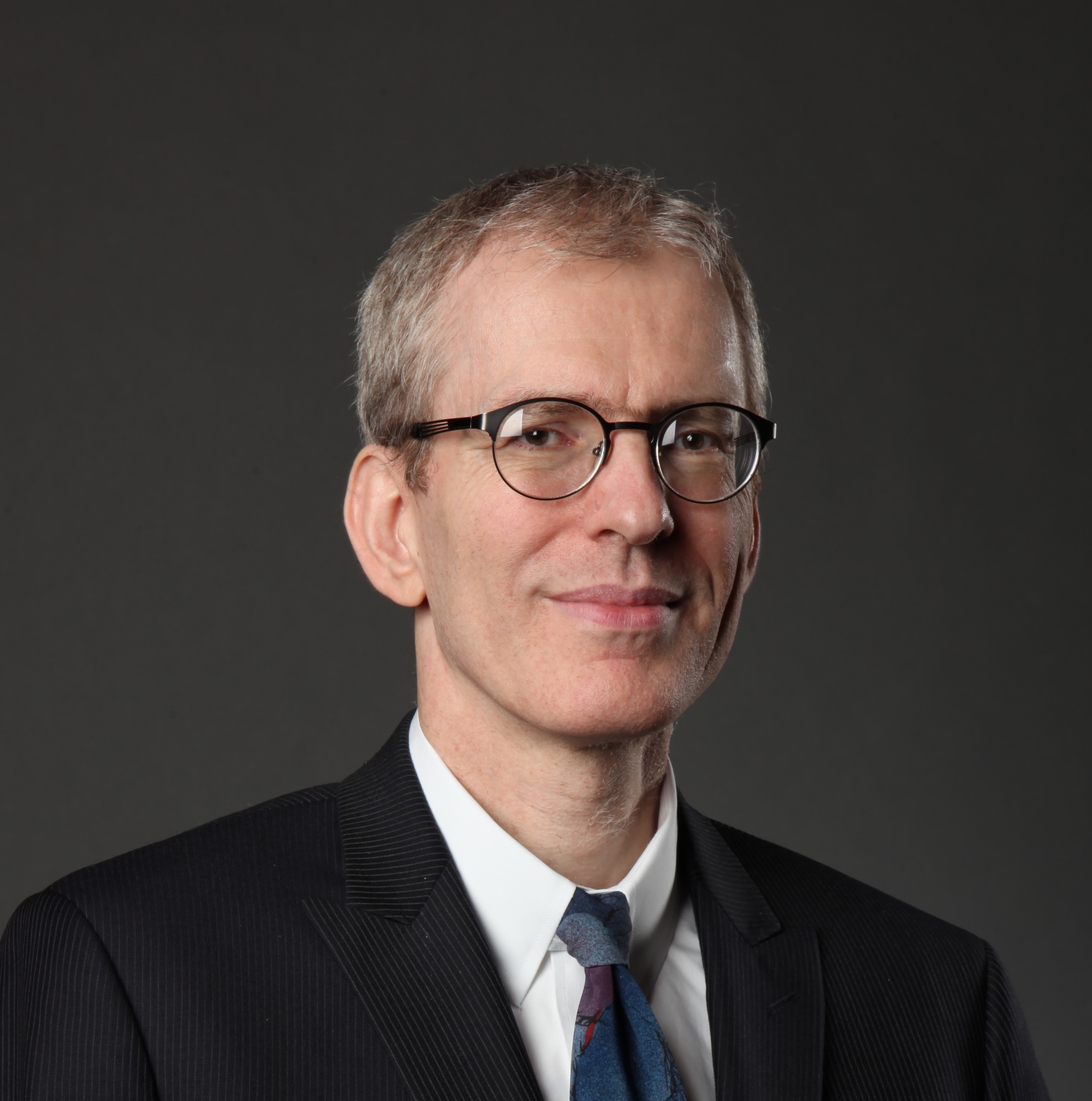}}]{Andreas F. Molisch} (F'05) received his degrees (Dipl.Ing. 1990, PhD 1994, Habilitation 1999) from the Technical University Vienna, Austria. He spent the next 10 years in industry, at FTW, AT\&T (Bell) Laboratories, and Mitsubishi Electric Research Labs (where he rose to Chief Wireless Standards Architect). In 2009 he joined the University of Southern California (USC) in Los Angeles, CA, as Professor, and founded the Wireless Devices and Systems (WiDeS) group. In 2017, he was appointed to the Solomon Golomb – Andrew and Erna Viterbi Chair.
 
His research interests revolve around wireless propagation channels, wireless systems design, and their interaction. Recently, his main interests have been wireless channel measurement and modeling for 5G and 6G systems, joint communication-caching-computation, hybrid beamforming, UWB/TOA based localization, novel modulation/multiple access methods, and machine learning for wireless systems. Overall, he has published 5 books (among them the textbook “Wireless Communications”, third edition in 2023), 22 book chapters, >320 journal papers, and >430 conference papers. He is also the inventor of 80 patents, and co-author of some 70 standards contributions. His work has been cited more than 79,000 times, his h-index is 118, and he is a Clarivate Highly Cited Researcher.
 
Dr. Molisch has been an Editor of a number of journals and special issues, General Chair, Technical Program Committee Chair, or Symposium Chair of multiple international conferences, as well as Chairperson of various international standardization groups. He is a Fellow of the National Academy of Inventors, AAAS, IEEE, IET, URSI, and AAIA, an IEEE Distinguished Lecturer, and a member of the Austrian Academy of Sciences. He has received numerous awards, among them the IET Achievement Medal, the Technical Achievement Awards of IEEE Vehicular Technology Society (Evans Avant-Garde Award) and the IEEE Communications Society (Edwin Howard Armstrong Award), and the Technical Field Award of the IEEE for Communications (Eric Sumner Award).
\end{IEEEbiography}

\end{document}